\documentclass[aps,pra,10pt,twocolumn,a4paper,nofootinbib,preprintnumbers]{revtex4-2} 

\usepackage{CJK}
\usepackage{amsmath,amsfonts,amssymb,amsthm}
\usepackage{eqnarray}
\usepackage{empheq}
\usepackage{float}
\usepackage{booktabs}

\usepackage{mathcomp} 
\usepackage{srcltx}
\usepackage[normalem]{ulem} 
\usepackage{graphics,graphicx}
\usepackage{dcolumn}
\usepackage{bm}
\usepackage{hyperref}
\usepackage{url}
\usepackage[mathlines]{lineno}
\usepackage{siunitx} 
\usepackage[dvipsnames]{xcolor} 
\usepackage{verbatim}
\usepackage{upgreek} 
\usepackage{color, colortbl}
\usepackage{multirow, tabularx, ragged2e, booktabs, makecell} 
\usepackage{fancyhdr}

\usepackage{array} 
\newcolumntype{L}[1]{>{\raggedright\let\newline\\\arraybackslash\hspace{0pt}}m{#1}}
\newcolumntype{C}[1]{>{\centering\let\newline\\\arraybackslash\hspace{0pt}}m{#1}}
\newcolumntype{R}[1]{>{\raggedleft\let\newline\\\arraybackslash\hspace{0pt}}m{#1}}

  \newcommand{\final}[1]{\textcolor{black}{#1}}

 \newcommand{\mm}[1]{\textcolor{black}{#1}}

\usepackage{color,soul}

\def\({\left(}
\def\){\right)}
\def\[{\left[}
\def\]{\right]}
\newcommand{\beq}{\begin{equation}}
\newcommand{\eeq}{\end{equation}}
\newcommand{\bea}{\begin{eqnarray}}
\newcommand{\eea}{\end{eqnarray}}

\definecolor{LightGray}{gray}{0.9}

\graphicspath{{./figures/}}

\usepackage{xr}

\begin{document}

\title{A Reconfigurable Multilayer Quantum Key Distribution Network over Existing Metropolitan Fibre}

\author{
Mariella Minder$^{\ast}$,
Andreas Siakolas,
Elizabeth Pasatembou,
Stylianos Mavrikos,
Stephanos Yerolatsitis,
Konstantinos Katzis
\& Kyriacos Kalli$^{\ast\ast}$
}

\affiliation{
\small
PhOSLab, Department of Electrical Engineering, Computer Engineering and Informatics, Cyprus University of Technology, Limassol 3036, Cyprus\\
$^{\ast}$mariella.minder@cut.ac.cy\\
$^{\ast\ast}$kyriacos.kalli@cut.ac.cy\\
}

\begin{abstract}
\noindent
Scaling quantum networks requires architectures extending end-to-end service reachability despite constrained fibre and equipment.
Critically, in brownfield deployments, the quantum network must be engineered around classical telecommunications infrastructure and the operating capabilities of quantum key distribution (QKD) technology.
To address this, we demonstrate a seven-node, multilayer QKD network deployed over existing metropolitan fibre.
The implementation compares spectral management techniques to accommodate inherited constraints across its layers.
It combines a trusted-node ring with a three-node subnetwork interconnected through reconfigurable optical switching, extending physical connectivity and service reachability without an additional transmitter.
A meshed key management layer maps endpoint requests onto trusted-relay paths and supplies Layer~1, Layer~3, and one-time-pad applications.
Over 73 days, eight link configurations generated 119.4~Gbit of secret key material with 99.1\% link availability.
Randomised end-to-end requests revealed indirect cross-layer coupling between key consumption, stored-key state and quantum-layer reconfiguration, demonstrating resource sharing through reconfigurability, spectral multiplexing and logical key-management abstraction.

\end{abstract}

\maketitle

\section*{Introduction}

Quantum key distribution (QKD) was introduced through the BB84 protocol, which showed that authenticated distant users can establish shared secret keys with security derived from quantum measurement rather than computational hardness assumptions~\cite{Bennett1984}.
Since then, QKD has progressed from laboratory demonstrations to deployed fibre and free-space links, and now represents the most mature quantum communication technology for near-term secure-network applications~\cite{Gisin2002,Scarani2009,Lo2014}.
Early field trials established the feasibility of QKD outside controlled laboratory environments, while subsequent metropolitan networks demonstrated trusted-node operation, multi-user key distribution, and long-term stability over deployed optical fibre~\cite{Elliot_2005,Peev_2009,Xu2009,Chapuran2009,Sasaki_2011,Stucki_2011}.

These early QKD networks established the feasibility, performance, and stability of multi-node quantum communication under field conditions.
Active optical switching was already explored as a means of sharing QKD resources and modifying quantum connectivity~\cite{Ma2007,Bogdanski2009,Poppe2007}, while hierarchical metropolitan networks combined optical switching, trusted relaying and application delivery over commercial telecommunications fibre~\cite{Xu2009}.
Deployments such as the DARPA Quantum Network~\cite{Elliot_2005}, SECOQC~\cite{Peev_2009}, the Tokyo QKD network~\cite{Sasaki_2011}, and SwissQuantum~\cite{Stucki_2011} were essential in demonstrating multi-node QKD, trusted relaying, and continuous operation, although many remained QKD-centric deployments in which the optical architecture was selected primarily around the requirements of the quantum links.

Later demonstrations moved closer to operational conditions by increasing network scale, improving long-term stability, sharing expensive quantum resources, and integrating QKD with conventional telecommunications infrastructure.
Resource-efficient architectures included wavelength-saving metropolitan networks~\cite{Wang2010Wavelength} and multi-user quantum access networks in which detection resources were shared among multiple users~\cite{Frohlich2013}.
The co-existence of QKD with classical telecommunications channels was also demonstrated using wavelength-division multiplexing over shared fibre with commercial QKD systems~\cite{Eraerds2010}, and high-bit-rate data services~\cite{Patel2012,Mao2018,Dynes2019}.

Software-defined-network (SDN)-controlled QKD networking and reconfiguration were subsequently demonstrated in metropolitan field trials~\cite{Ou2018,Tessinari2019,Lopez2020}, while a 46-node metropolitan network combined trusted relays, optical switches, systematic key management and application modules, including key-store-aware scheduling of quantum connections~\cite{Chen2021Hefei}.
European deployments further advanced integration with telecommunications infrastructure: the OpenQKD testbeds investigated field-installed QKD and application integration~\cite{Braun2021OpenQKD}, and MadQCI demonstrated heterogeneous QKD technologies, dynamically switched quantum connections, standards-based SDN control and multiple security-service layers within production telecommunications facilities~\cite{Martin2024}.
At larger scales, national and intercity QKD infrastructures, including terrestrial trusted-node backbones, satellite-assisted links, and undersea channels, have explored quantum-secure communications beyond metropolitan distances~\cite{Chen_2021,Chen_2025,Liao2017,AmiesKing2023}.

As QKD networks' priorities shift towards integration with operational telecommunications infrastructure, the central problem is no longer only the secure key rate of an individual link.
It is also the service reachability obtained from a finite set of fibres, wavelengths, and QKD transceivers; the ability to modify physical connectivity as demand changes; and the capacity of the control and key management layers to
translate heterogeneous physical links into end-to-end services.
This problem is particularly acute in brownfield deployments, where fibre routes, intermediate distribution points, and available resources are largely predetermined by the classical network, while current commercial QKD systems remain less flexible than mature telecommunications equipment and remain constrained by finite optical loss budgets, wavelength requirements, and limited configurability.
The resulting challenge is to integrate and reconfigure the available quantum resources within these practical constraints while still providing useful end-to-end services.
Resource-aware routing, quality-of-service metrics and software-defined key management have been developed to address aspects of this problem~\cite{Mehic2020Networking,Cho2021}.
Deployed networks have also demonstrated dynamic optical connectivity~\cite{Martin2024} and key-store-aware resource allocation~\cite{Chen2021Hefei}.
However, the operational trade-offs associated with deploying and reconfiguring constrained QKD resources remain comparatively under-characterised, particularly the relationship between physical-resource reuse, service reachability, optical integration constraints, reconfiguration overhead, stored-key dynamics, and end-to-end service behaviour.

In this work, we address these challenges in a seven-node metropolitan QKD network deployed over a pre-existing fibre plant in Nicosia, Cyprus, serving governmental entities.
The deployment comprises two use cases (UCs) with distinct quantum-layer topologies: UC\textsubscript{1}, a four-node trusted-node ring, and UC\textsubscript{2}, a three-node reconfigurable subnetwork, with optical switching providing interconnection between them.
A logically meshed KMS provides direct and trusted-relay key delivery, while an SDN-based control and telemetry layer supports network monitoring and reconfiguration.
The reconfigurable interconnection extends logical reachability from 9 to 21 endpoint pairs through trusted relaying.
The resulting multi-layer infrastructure supplies quantum-generated keys to Layer~1 encryption, Layer~3 encryption and one-time-pad applications.

We evaluate both the benefits and operational costs of this architecture under field conditions.
Over 73 days of continuous monitoring, the eight link configurations generated 119.4~Gbit of secret key material with a mean reported link availability of 99.1\%.
Controlled reconfiguration showed that transmitter reuse incurs an additional $17\pm5$~min before fresh key material becomes available after switching.
For UC\textsubscript{1}, comparison of two optical implementations and sequential link activation quantify the trade-off associated with spectral filtering and verify simultaneous operation of the ring links.
Finally, randomised end-to-end requests probe trusted-relay routing and stored-key depletion under changing demand and physical connectivity.
Together, these results quantify the reachability gained through physical-resource reuse alongside its optical, reconfiguration and key-management consequences in a deployed metropolitan QKD network.
The deployment also constitutes, to our knowledge, the first multi-node QKD network demonstrated in Cyprus.

\section*{Results}

\begin{figure*}[t]
	\centering
	\includegraphics[width=1\textwidth]{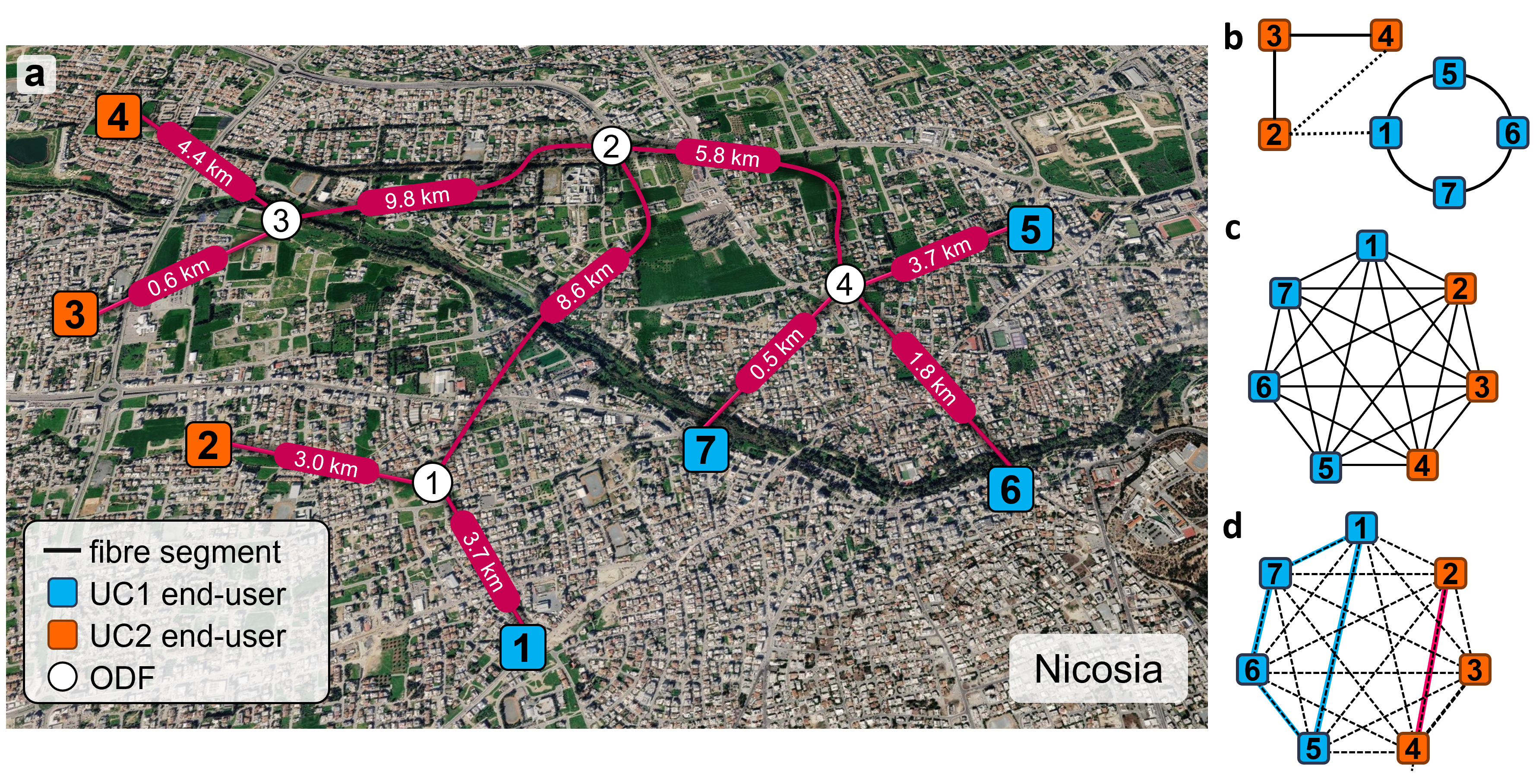}
	\caption{\textbf{Multi-layer architecture of the deployed QKDN.} 
		\textbf{a,} Layout of the deployed fibre infrastructure in Nicosia, showing the end-user nodes of use-case~1 (UC1) and use-case~2 (UC2), intermediate optical distribution frame (ODF) sites, and fibre segments with their measured lengths.
        Map credits: Imagery \textcopyright2026 Google, Airbus, Landsat / Copernicus, Maxar Technologies, CNES/Airbus.
        \textbf{b,} Physical quantum-layer topology, including the UC1 ring, the UC2 links, and the alternative N2--N1 and N2--N4 optical-switch configurations.
        \textbf{c,} Logical key management system topology, providing any-to-any key delivery between nodes through trusted-node relaying.
        \textbf{d,} Application-layer service topology, showing the deployed secure-service links supported by QKD-derived key material.
        Dashed lines indicate the endpoint pairs accessible to the on-demand OTP file-encryption application.
        Layer~1 and Layer~3 encryption services are the blue and red lines, respectively.
		}
	\label{fig:map}
\end{figure*}

\textbf{Network architecture and service reachability.}
The deployed network consists of two functional classes of governmental end-users (nodes), defining two UCs with distinct quantum-layer topologies (Fig.~\ref{fig:map}a, b).
UC\textsubscript{1} comprises nodes N1, N5, N6 and N7 arranged as a four-node trusted-node ring, while UC\textsubscript{2} comprises nodes N2, N3 and N4, with permanent links N2–N3 and N3-N4.
The network is deployed over 42~km of pre-existing metropolitan fibre, with a total optical path length of 108~km and four intermediate optical distribution frames (ODFs).
In each path segment, one dark fibre is allocated to the quantum channels and a second fibre to the classical plane, which can also support conventional telecommunications traffic.
The lengths and losses of each segment were characterised using optical time-domain reflectometry, see Supplementary Information (SI).

At ODF1 coordinated switching of the quantum and synchronisation paths connects the transmitter at N2 either to the receiver at N4 or to the receiver at N1.
In the N2--N4 state, the direct N2--N4 link completes the three-node UC\textsubscript{2} mesh and the two use cases operate independently.
In the N2--N1 state, the direct N2--N4 link is replaced by an inter-subnetwork connection, producing a connected seven-node trusted-relay topology.
The N2--N4 and N2--N1 links are therefore mutually exclusive and share the same QKD transmitter.

This reconfiguration changes the logical reachability of the network.
With N2--N4 selected, the disconnected four-node and three-node subnetworks support $6+3=9$ within-subnetwork endpoint pairs.
Selecting N2--N1 instead connects all seven nodes, making all 21 endpoint pairs logically reachable through trusted relaying, subject to sufficient stored key material and the trusted-node security assumptions.
The shared transmitter therefore increases logical endpoint-pair reachability by 133\% relative to the disconnected configuration, without provisioning an additional QKD transmitter.

\begin{figure*}[th]
    \centering
    \includegraphics[
        width=\textwidth
    ]{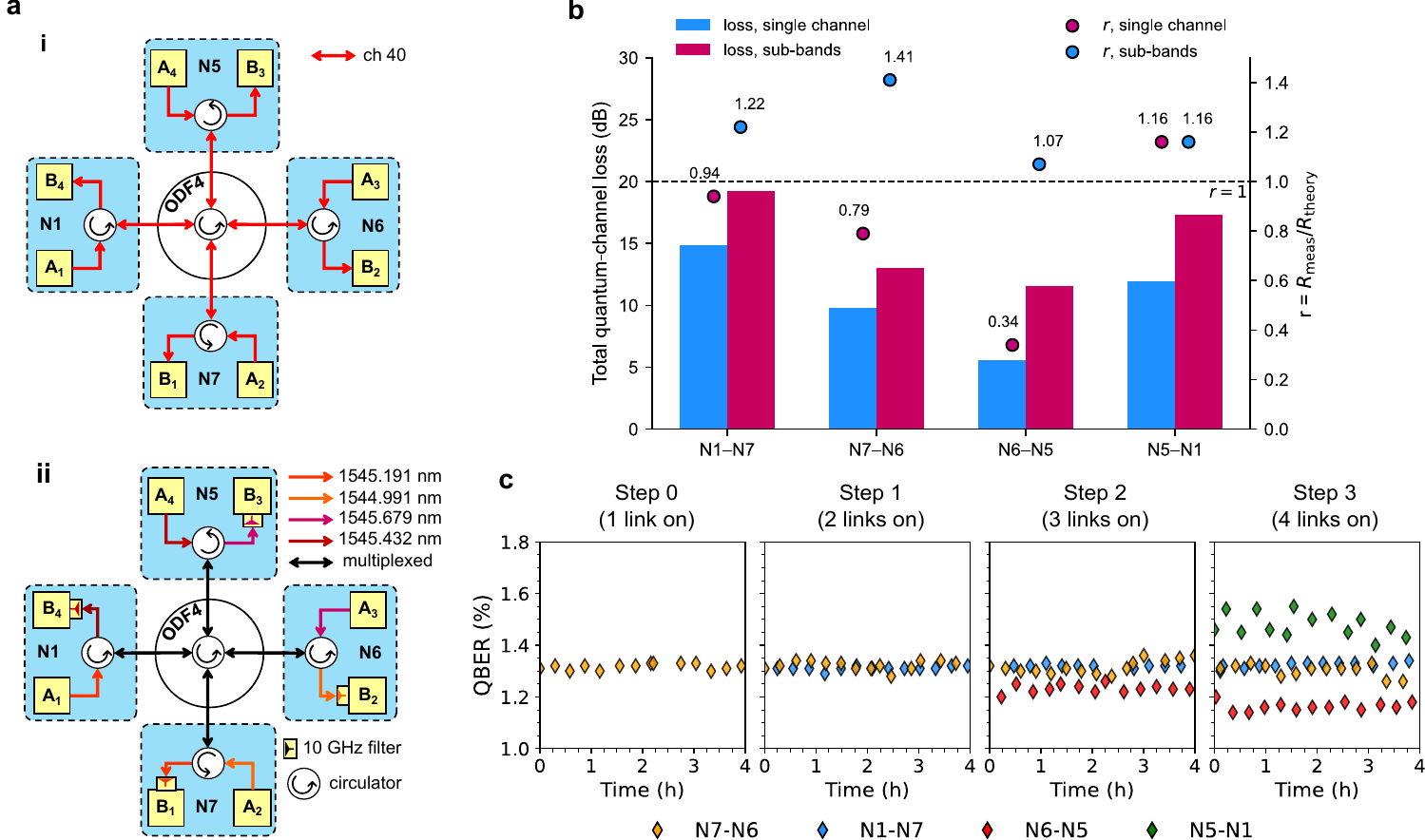}
    \caption{\
    \textbf{Optical implementation of the UC\textsubscript{1} shared-wavelength architecture of the quantum channel.}
    \textbf{a}, UC\textsubscript{1} quantum-layer configuration before (\textbf{i}) and after (\textbf{ii}) introducing intra-channel spectral separation.
    In the initial\mm{, single-wavelength,} configuration, the four QKD links share the nominal \mm{100~GHz} ITU-T DWDM channel~40; in the \mm{revised, }sub-band, configuration, the transmitters are wavelength-offset within channel~40 (ch~40) and isolated using 10~GHz optical filters.
    \textbf{b}, Total quantum channel loss (bars, left axis) and loss-normalised secret-key rate, $r=R_{\mathrm{meas}}/R_{\mathrm{theory}}$ (markers, right axis), for the four UC\textsubscript{1} links in the single-\mm{wavelength} and sub-band configurations.
    \textbf{c}, Sequential activation of the four sub-band-separated links, from one active link in Step 0 to simultaneous operation of all four links in Step 3; colours identify the individual UC1 links.
    }
    \label{fig:spectralMethods}
\end{figure*}

The network incorporates additional KMS and application layers to enable end-to-end key management and secure service delivery across the interconnected sites (Fig.~\ref{fig:map}c,d). 
In contrast to the quantum layer, whose connectivity is constrained by the available QKD links and the optical-switch state, the KMS layer accepts logical key requests between any pair of nodes and maps them onto direct or trusted-relay paths, provided that sufficient stored key material is available on each required physical link.
This logical reachability is enabled by regenerating and switching the classical KMS connectivity at the intermediate ODF sites, thereby creating a logically meshed key-management layer over the fixed fibre plant.
The same key-delivery layer supplies QKD-derived keys to Layer~1 encryption, deployed between all UC\textsubscript{1} nodes, Layer~3 IP encryption deployed on the N2--N4 link, and an on-demand one-time-pad application available to all users.

\textbf{Optical integration over existing fibre.}
Each route segment used separate fibres for the quantum and classical planes. The quantum channel fibres carried wavelength-multiplexed QKD signals, whereas synchronisation, KMS and application traffic were wavelength-multiplexed over the second, classical fibre.
\mm{The inherited fibre topology required different wavelength-management strategies in the two use cases.}

In UC\textsubscript{1}, several links propagated through shared ring segments using circulator-defined paths.
This was implemented using two different techniques.
Hereafter, channel numbers refer to the 100~GHz ITU-T dense wavelength division multiplexing (DWDM) grid in the C-band.
In the initial implementation, described in ~\cite{Minder2025SPIE}, all four QKD carriers operated at the nominal DWDM channel~40  ($194.0~\mathrm{THz}$;
$\lambda \approx 1545.32~\mathrm{nm}$), Fig.~\ref{fig:spectralMethods}a,i, while the corresponding synchronisation carriers, transmitted over the classical fibre, similarly shared a common wavelength channel.
These signals were routed through the ring using optical circulators.
\final{However, finite circulator isolation permitted leakage between co-located transmitters and receivers, with the resulting back-reflections observed to affect link performance.}
The UC\textsubscript{1} wavelength-management architecture was therefore redesigned using different approaches for the quantum and synchronisation signals.
The nominally same-channel quantum carriers remained within channel~40 but were spectrally separated by temperature tuning and narrowband 10~GHz filtering (Fig.~\ref{fig:spectralMethods}a,ii), in order to avoid replacing the specialised quantum signal source.
Instead, the conventional synchronisation sources were modified to operate on distinct DWDM channels, providing wavelength separation between the corresponding synchronisation paths.

\begin{figure*}[th]
    \centering
    \includegraphics[
        width=\textwidth
    ]{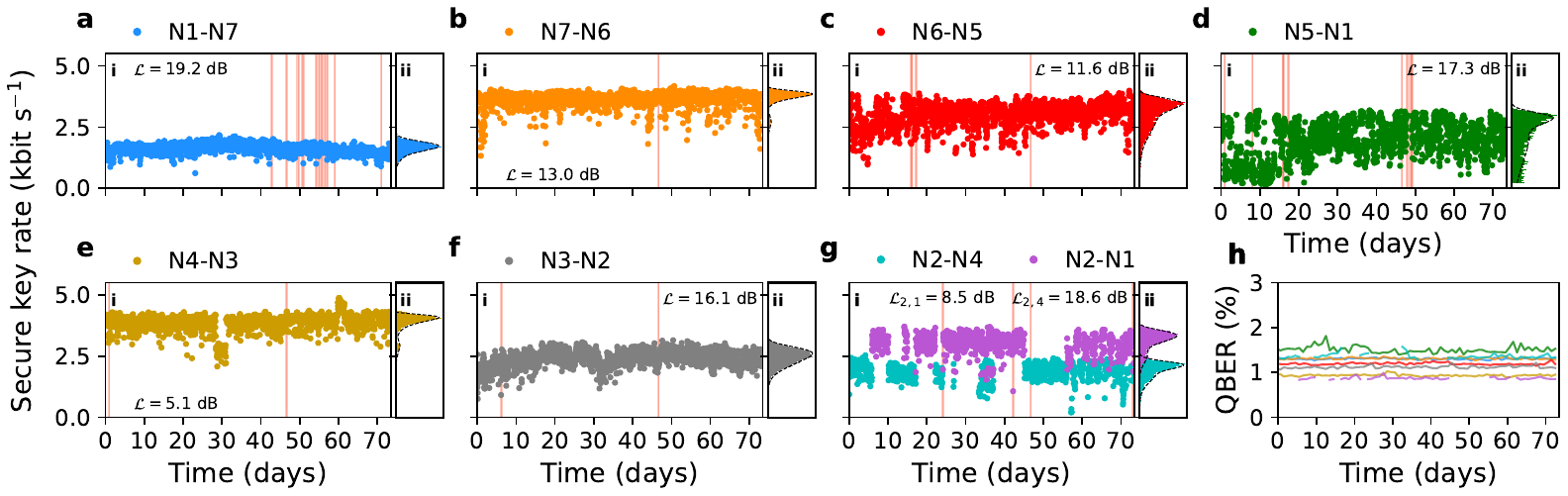}
    \caption{\
    \textbf{Long-term quantum-layer performance over 73 days.}
    \textbf{a--g}, Time-resolved secure key rate (SKR) for each deployed QKD link.
    In each panel, \textbf{i} shows the logged SKR values, with the mean measured channel loss denoted by $\mathcal{{L}}$, and red vertical lines marking link-unavailability events.
    \textbf{ii} shows the corresponding SKR distribution.
    \mm{For visualisation, the black curves show two-component reflected log-normal fits to the SKR distributions.}
    \textbf{h}, Temporal evolution of the quantum bit error rate (QBER) for all QKD links; colours correspond to the links shown in \textbf{a--g}.}
    \label{fig:timeseries}
\end{figure*}

\mm{Fig.~\ref{fig:spectralMethods}b compares and summarises the field performance of the two approaches.}
The additional quantum signal filtering and routing increased the mean total quantum channel loss across the four UC\textsubscript{1} links from 10.6 to 15.3~dB.
Even so, the mean secure key rate (SKR) increased 2.2 to 2.6~kbit~s$^{-1}$, while the mean quantum bit error rate (QBER) decreased from 1.48\% to 1.33\%.
To account for \mm{the difference in} channel loss \mm{between implementations}, we normalised the averaged measured SKR of each UC\textsubscript{1} link, $R_{\mathrm{meas}}$, to the corresponding loss-dependent model prediction, $R_{\mathrm{meas}}/R_{\mathrm{theory}}(\mathcal{L})$.
For the initial implementation, $R_{\mathrm{meas}}$ was obtained from a 30-day monitoring dataset of the network described in Ref.~\cite{Minder2025SPIE}, whereas for the revised implementation it was obtained from the 73-day monitoring dataset reported here; in each case, $R_{\mathrm{theory}}$ was evaluated at the measured loss of that implementation.
This ratio increased for three of the four links and remained unchanged for one, changing from 0.94, 0.79, 0.34 and 1.16 in the previous implementation to 1.22, 1.41, 1.07 and 1.16 in the revised implementation for N1--N7, N7--N6, N6--N5 and N5--N1, respectively (Fig.~\ref{fig:spectralMethods}b).
The arithmetic mean of the four per-link ratios consequently increased from 0.81 to 1.22, corresponding to an approximately 50\% relative increase.
The improved key-generation performance despite the additional optical loss is consistent with the combined wavelength management redesign mitigating impairments present in the original configuration.

To test the effects of simultaneous operation of the \mm{intra-channel wavelength-separated signals} directly, the four UC\textsubscript{1} links of \mm{the revised implementation} were activated sequentially while monitoring the already-operating links (Fig.~\ref{fig:spectralMethods}c).
No systematic degradation was observed as additional carriers were enabled, indicating that the four sub-band-separated quantum carriers could operate simultaneously within the shared nominal ITU channel.
Importantly, the specialised quantum optical sources themselves were not modified.
Their wavelengths were adjusted using the available laser temperature tuning, and separated using external narrowband filtering, while the comparatively flexible synchronisation sources were modified to operate on distinct 100~GHz DWDM channels.
The revised architecture therefore introduced the required wavelength flexibility without altering the specialised quantum-source hardware.

In UC\textsubscript{2}, the quantum links instead occupied distinct DWDM channels.
Wavelength-selective routing at the intermediate ODF sites provided the fixed connections, while coordinated optical switching of the quantum and synchronisation paths enabled the alternative N2--N4 and N2--N1 configurations.
The two subnetworks therefore employed complementary wavelength management strategies of the quantum signals: intra-channel spectral separation in UC\textsubscript{1} and conventional DWDM separation in UC\textsubscript{2}.

Across the deployed quantum paths of both use-cases, the passive optical infrastructure introduced between 3 and 14~dB of loss in addition to the bare-fibre attenuation, arising from circulators, narrowband filters, wavelength-routing components, connectors and, for the reconfigurable path, the optical switch.
The complete optical implementation and the corresponding fibre and channel losses are provided in Supplementary Fig.~S1 and Supplementary Table~S3.

\begin{figure}[t]
	 \centering
    \includegraphics[
        width=\columnwidth
    ]{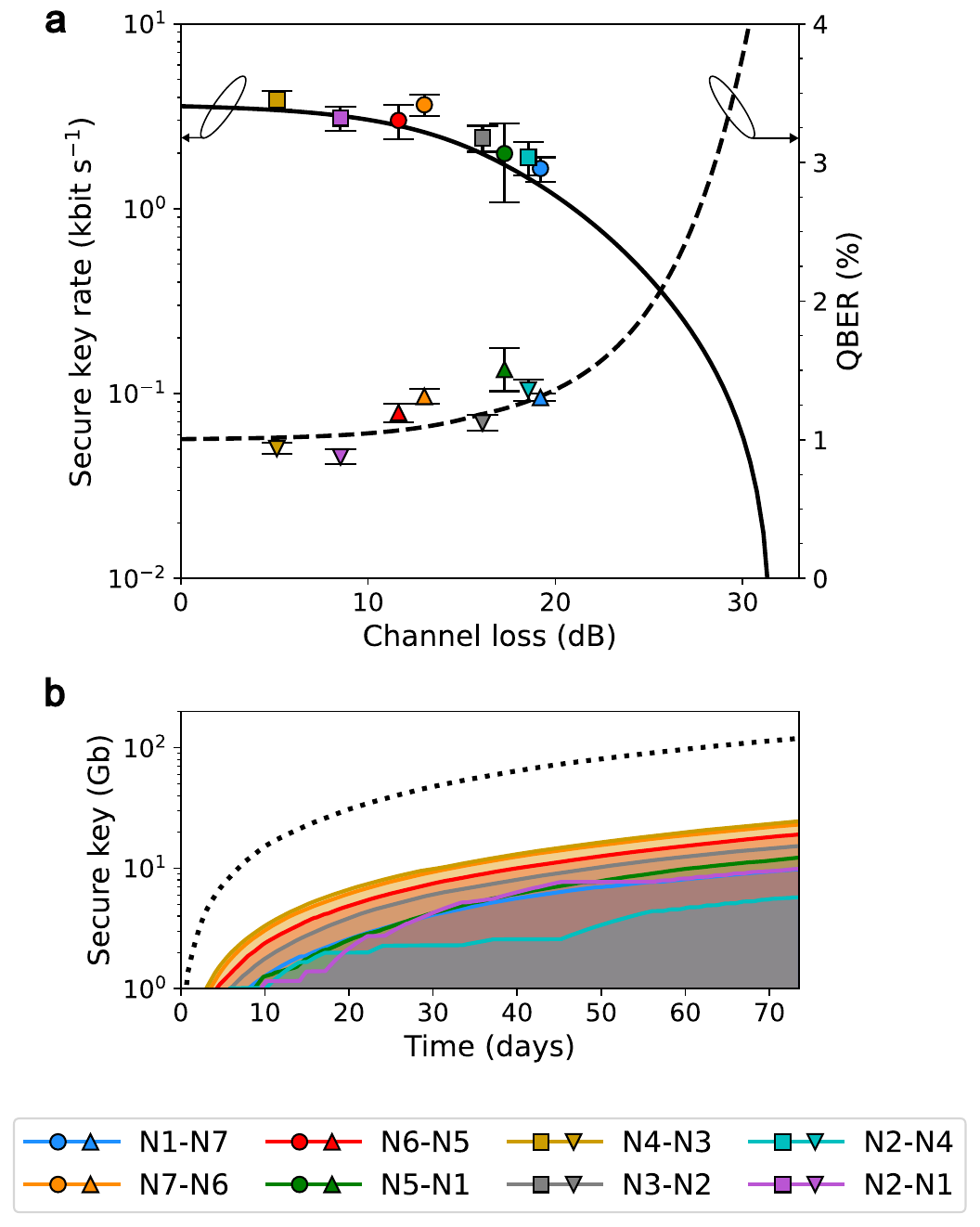}
	\caption{\textbf{Channel-loss-dependent QKD performance and cumulative secret-key generation over 73 days.}
    \textbf{a}, Mean secure key rate (SKR; circles for UC\textsubscript{1}, squares for UC\textsubscript{2}, left axis) and QBER (triangle-up for UC\textsubscript{1}, triangle-down for UC\textsubscript{2} right axis) measured for each QKD link as a function of total quantum channel loss.
    Error bars indicate one standard deviation over the monitoring period.
    The solid and dashed black curves show the corresponding finite-key model predictions for the SKR and QBER, respectively.
    \textbf{b}, Cumulative secret-key material generated by each QKD link during the 73-day monitoring period.
    The dotted black curve represents the network-wide cumulative total.
    Colours identify the links as specified in the legend.}
	\label{fig:theory}
\end{figure}

\medskip

\textbf{Long-term operational performance.}
The 73-day monitoring period evaluates the sustained operation of the complete optical integration rather than isolated point-to-point QKD devices, including wavelength multiplexing, circulator-defined paths, narrowband filtering, intermediate ODF routing, and the switched connection.
Fig.~\ref{fig:timeseries}a-g present the time-resolved SKR and corresponding active-state SKR distribution for each of the eight link configurations.
The temporal mean SKR of the individual link configurations ranged from 1.7 to 3.9 kbit~s$^{-1}$.
The arithmetic mean of the eight per-link temporal means was 2.7 kbit~s$^{-1}$, with a standard deviation across link configurations of 0.8 kbit~s$^{-1}$.
\mm{Although the temporal SKR distributions differed between links, several exhibited bimodal distributions or pronounced low-rate populations, indicating recurrent operation at distinct key-generation levels rather than simple transitions between normal operation and outage.
These distributions were therefore represented using two-component reflected log-normal fits as a descriptive model of the observed operating regimes, without assigning a specific physical origin to the individual components.
Despite this variability, secret-key generation was sustained with a mean reported link availability of 99.1\%.}
Performance and availability metrics are summarised in Table~\ref{tab:performance}.

\begin{table}[htbp]
    \centering
    \caption{Average QKD-link performance and availability.}
    Per-link uncertainties denote temporal standard deviations over the monitoring period.
    \label{tab:performance}
    \begin{tabular}{cccc}
        \toprule
        Link &
        \makecell{$\mu_{\mathrm{SKR}}\pm\sigma_{\mathrm{SKR}}$ \\[2pt]
        (kbit~s$^{-1}$)} & \makecell{$\mu_{\mathrm{QBER}}\pm\sigma_{\mathrm{QBER}}$ \\[2pt] (\%)} &
        \makecell{Link \\availability \\[2pt] (\%)} \\[2pt]
        \midrule
        N1--N7 & $1.7 \pm 0.3$ & $1.30 \pm 0.03$ & 96.6 \\[2pt]
        N7--N6 & $3.7 \pm 0.5$ & $1.31 \pm 0.05$ & 99.9 \\[2pt]
        N6--N5 & $3.0 \pm 0.6$ & $1.19 \pm 0.07$ & 99.5 \\[2pt]
        N5--N1 & $2.0 \pm 0.9$ & $1.50 \pm 0.20$ & 97.7 \\[2pt]
        N4--N3 & $3.9 \pm 0.4$ & $0.94 \pm 0.04$ & 99.9 \\[2pt]
        N3--N2 & $2.4 \pm 0.4$ & $1.12 \pm 0.06$ & 99.9 \\[2pt]
        N2--N4 & $1.9 \pm 0.4$ & $1.36 \pm 0.08$ & 99.7 \\[2pt]
        N2--N1 & $3.1 \pm 0.5$ & $0.88 \pm 0.05$ & 99.2 \\[2pt]
        \midrule
        Mean across\\links& $2.7\pm0.8$ & $1.2\pm0.2$ & $99.1 \pm 1$ \\
        \bottomrule
    \end{tabular}
\end{table}

Fig.~\ref{fig:timeseries}h shows the temporal evolution of the QBER.
The arithmetic mean of the eight per-link temporal mean QBER values was 1.2\%, with a standard deviation across link configurations of 0.2\%.
QBER remained within a narrow operating range during key-generation periods, with individual-link temporal means between 0.88\% and 1.50\%.
The mean SKR and QBER \mm{of each link} are compared with the corresponding \mm{finite-key model predictions as a function of total quantum channel loss} in Fig.~\ref{fig:theory}a.
The agreement between measurement and model shows that the long-term \mm{mean performance} of the deployed links is consistent with the expected loss-dependent performance of the implemented QKD protocol.

Finally, Fig.~\ref{fig:theory}b presents the cumulative secret key material generated by each link over the 73-day monitoring period.
Summing the generated material across all key pools gives an aggregate total of $119.4~\mathrm{Gbit}$ of link-local secret key material during the monitoring period.
This aggregate is not equivalent to the volume of end-to-end key available between arbitrary endpoint pairs, because trusted-relay delivery consumes key material on every physical hop of a route.
Changes in the slopes of the cumulative-key traces reflect variations in key-generation rate, availability and, for links N2--N1 and N2--N4, optical switching events.
\begin{figure*}[t!]
	\centering
    \includegraphics[width=\textwidth]{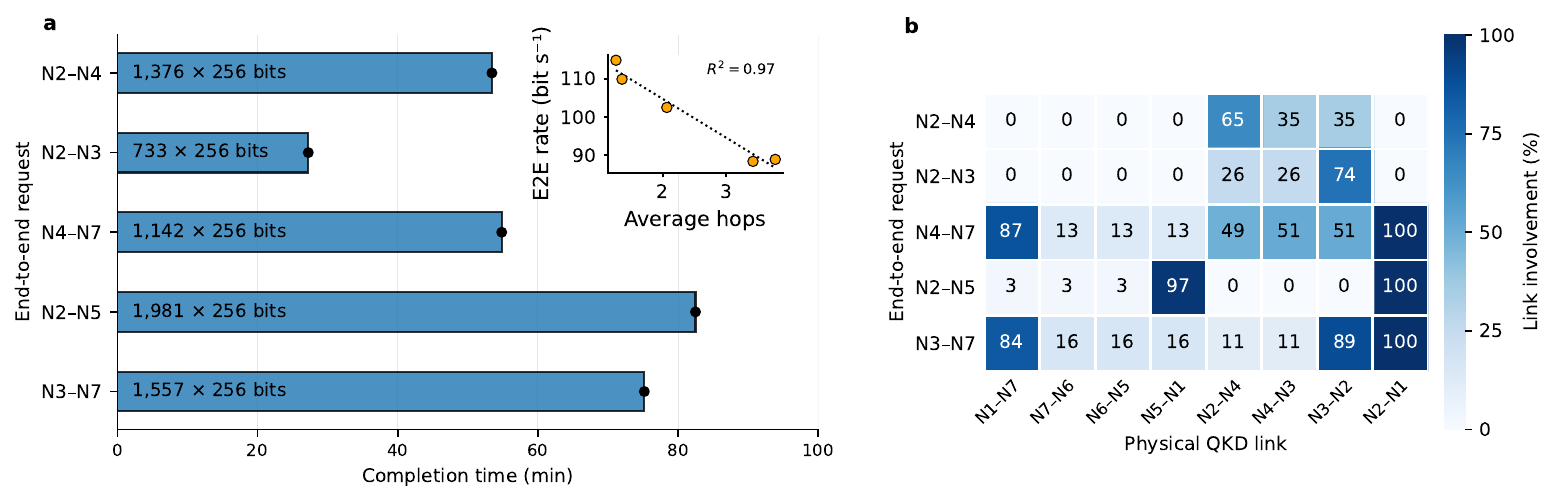}
	\caption{
    \textbf{End-to-end key delivery and physical-link involvement.}
    \textbf{a}, Bars show the time required to deliver the specified number of 256-bit keys, with the corresponding effective end-to-end delivery rate indicated at the end of each bar.
    The inset plot shows the effective end-to-end key-delivery rate as a function of the average number of physical QKD hops used per route-resolved delivered key.
    The dashed line is a linear fit to the five request intervals; $R^2=0.97$ is shown as a descriptive measure of the observed association.
    \textbf{b,} Physical-link involvement for each end-to-end request. Each cell gives the percentage of route-resolved delivered keys whose reconstructed trusted-relay path traversed the corresponding physical QKD link.
    }
	\label{fig:randomTest}
\end{figure*}

\textbf{Reconfiguration overhead.}
The N2--N4 and N2--N1 configurations share the transmitter at N2 and were therefore operated alternately.
During days 35--37, the optical switch was commanded to alternate between the two configurations every 2~h in a controlled experiment designed to quantify the operational overhead associated with transmitter reuse.
The time from each switch command to the first fresh secret key produced by the newly selected link was compared with the corresponding key-generation interval during periods without switching.
The analysis included 37 reconfiguration intervals and 253 static intervals and yielded an additional latency of $17\pm 5~\mathrm{min}$.
This latency includes the actuation time of the optical switch itself, the subsequent QKD-system calibration, and the start-up time required before fresh secret-key material becomes available on the newly selected link.
The measurement therefore quantifies the finite reconfiguration cost incurred when a single QKD transmitter is reused between the local N2--N4 connection and the inter-subnetwork N2--N1 connection.
A zoomed view of the resulting SKR traces, as well as the corresponding key generation time are provided in Supplementary Fig.~S2.

\textbf{End-to-end key delivery under dynamic demand and physical reconfiguration.}
To examine how the multi-layer architecture translates changing application demand into use of the underlying QKD resources, we performed a randomised target-volume end-to-end key-delivery experiment.
During each request interval, a source--destination pair and target key volume were selected, and the KMS delivered sequential 256-bit end-to-end keys until the target was reached before proceeding to the next request interval (Fig.~\ref{fig:randomTest}).
Five endpoint pairs were exercised over approximately 5~h, with target volumes randomly selected between 500 and 2001 256-bit keys.
All five target volumes were delivered successfully, with completion times ranging from approximately 27 to 82~min and effective end-to-end key-delivery rates between 88.4 and 114.8~bit~s$^{-1}$ (Fig.~\ref{fig:randomTest}a).
The experiment was performed while the deployed Layer~1 and Layer~3 encryption services remained active and continued to consume QKD-derived key material from their corresponding link-local key pools.

For each request, the QSDN controller selected a direct or trusted-relay path using a Dijkstra-based routing algorithm~\cite{Dijkstra1959}.
The commercial routing metric incorporates several physical-link parameters, including stored key material, quantum channel loss and SKR.
Because the precise cost function and parameter weights are proprietary, the routing algorithm was not independently reimplemented or optimised.

Delivery of each end-to-end key consumes the corresponding key material from every physical QKD-link key pool traversed by its trusted-relay path.
Physical-link involvement was therefore reconstructed from the observed link-pool decrements, constrained by the known physical topology, as described in Methods.
Figure~\ref{fig:randomTest}b shows, for each end-to-end request, the percentage of route-resolved delivered keys whose reconstructed path traversed each physical QKD link.

Across the five request intervals, the average reconstructed path length ranged from 1.26 to 3.78 physical QKD hops per route-resolved delivered key.
The effective end-to-end delivery rate showed a strong inverse association with average hop count, with a descriptive linear fit giving $R^2=0.97$ (Fig.~\ref{fig:randomTest}a, inset).
The two shortest-path requests, averaging 1.26--1.35 hops, achieved 109.9--114.8~bit~s$^{-1}$, whereas the two longest-path requests, averaging 3.43--3.78 hops, achieved 88.4--88.9~bit~s$^{-1}$.
Given the small number of request intervals and concurrent variation in route selection and stored-key state, this relationship is interpreted as an operational association rather than a causal scaling law.
Further details on the end-to-end key test are given in the Supplementary Information.

\mm{Importantly, the KMS operates on stored link-local key material, so an end-to-end trusted-relay request does not require every constituent QKD link to be generating at the time of the request, provided that sufficient key material is available on each required link.
This buffering is particularly relevant to the mutually exclusive N2--N4 and N2--N1 connections, for which optical reconfiguration determines which link can replenish its key pool at a given time.
Switch logs recorded reconfiguration during two of the five request intervals; following each switch, the newly selected QKD link underwent calibration before fresh key generation resumed, see Supplementary Fig.~S3e.
KMS routing and optical-switch control remained distinct—the former selected the logical trusted-relay path, whereas the latter determined which switched physical link generated fresh keys—but were indirectly coupled through the link-local key pools, which are depleted by end-to-end requests and replenished by QKD generation, with their availability and production--consumption balance informing reconfiguration.
The experiment therefore shows how stored-key buffering decouples instantaneous service delivery from instantaneous quantum-key generation, while preserving a feedback mechanism through which service demand can influence subsequent allocation of quantum resources.}

\section*{DISCUSSION}
The deployment demonstrates the practical integration of a complex QKD network stack within an existing metropolitan fibre infrastructure.
Seven governmental sites were interconnected through two quantum layer architectures, combining circulator-based optical routing, intra-channel spectral subdivision, DWDM, coordinated switching of quantum and synchronisation paths, trusted-node relaying, a logically meshed KMS, SDN-based control and telemetry, and key delivery to Layer~1, Layer~3 and one-time-pad applications.
Although these technologies and techniques have individually been demonstrated previously, their integration within a common brownfield network required them to operate within the same inherited fibre topology, loss budgets, wavelength constraints, and equipment limitations.
Its significance therefore lies not only in the performance of its individual links, but in sustained interoperability across optical, key-management, control and application layers under realistic network constraints.

The UC\textsubscript{1} redesign illustrates a broader challenge of brownfield QKD deployment: the optical \mm{characteristics} of commercial QKD equipment \mm{may} not \mm{align with the limitations} imposed by an inherited network architecture.
Here, the specialised quantum sources were preserved unchanged and accommodated using their available temperature tuning together with external intra-channel filtering, whereas the comparatively flexible synchronisation sources were modified to operate on distinct ITU-T channels.
This hybrid approach introduced wavelength flexibility where it was technically accessible while avoiding modification of the specialised quantum-source hardware.
Despite introducing approximately 5~dB of additional mean quantum channel loss, the adapted architecture exhibited improved key-generation performance, indicating that impairments beyond attenuation had limited the original configuration.
The result shows that additional passive loss, \mm{especially in short distance deployments,} can be an acceptable trade-off when it suppresses more consequential system-level impairments.
More generally, transferring wavelength flexibility to the surrounding optical infrastructure where direct modification or replacement of the quantum source is impractical provides a practical route for integrating wavelength-constrained QKD equipment into network architectures for which it was not originally configured.

A complementary challenge in QKDN deployments is the efficient use of the QKD hardware itself.
Dedicated QKD systems represent a substantial capital investment, making it impractical in many deployment scenarios to provision a separate system for every potential connection or network extension.
Physical resource sharing allows already deployed hardware to serve different connectivity requirements over time.
In the present network, reassigning a single transmitter increases the number of logically reachable endpoint pairs from 9 to 21, a 133\% increase.
This gain comes at the cost of mutually exclusive operation of the two alternative physical links and an additional 17~min before fresh key generation resumes following reconfiguration.
Even so, previously generated keys remain available in the link-local stores and can continue to support end-to-end delivery while a link is inactive or recalibrating.
For services whose key-consumption rates are small relative to the available stored-key capacity, the resulting trade-off therefore favours substantially improved hardware utilisation and logical reachability at the cost of delayed key replenishment rather than loss of service availability.

The practical value of such resource sharing depends critically on stored secret key.
In a trusted-relay QKD network, end-to-end delivery need not coincide with key generation on every physical link of the selected route.
Previously generated link-local key material can remain available to the KMS while the corresponding QKD connection is inactive, partially decoupling instantaneous physical generation from logical service.
This behaviour was observed for the mutually exclusive N2--N1 and N2--N4 connections, where stored key could continue to support trusted-relay delivery while the shared transmitter was assigned to the alternative connection.
At the same time, trusted relaying imposes a service-level resource cost because every end-to-end key consumes stored key material from each physical QKD link along its route.
Across the five request intervals, average reconstructed path lengths ranged from 1.26 to 3.78 hops, with the longer-path requests associated with lower effective end-to-end delivery rates.
Stored key therefore acts both as a buffer and as a finite network resource linking previous generation, application demand and future physical-resource allocation.

The 73-day monitoring period tests this integrated architecture as an operational system rather than as isolated demonstrations.
Across the eight physical-link configurations, the network generated 119.4~Gbit of link-local secret key material with a mean reported link availability of 99.1\%, showing that the added optical and control complexity remained compatible with sustained QKD operation.
These results complement other advanced metropolitan QKD deployments, including MadQCI~\cite{Martin2024}, and the recent Milan metropolitan
network~\cite{DeLazzari2026Milan}, which have demonstrated combinations of optical switching, trusted relaying, key-management layers and software-defined control in operational fibre environments.
The distinction of the present work is therefore not network size or the first use of these technologies, but their field integration under inherited infrastructure constraints together with quantitative characterisation of the resulting optical, temporal, reachability, stored-key, and end-to-end service trade-offs.

\mm{Future work will focus on reducing the optical overhead associated with network integration, for example through lower-loss routing and filtering architectures, while introducing quantum--classical coexistence over shared fibre to further reduce infrastructure requirements.
Extending the architecture beyond the metropolitan scale to longer inter-city links would test whether the same principles of spectral adaptation, resource sharing and stored-key management remain effective under tighter optical-loss and key-generation constraints. 
At the control layer, replacing or adapting the proprietary KMS and SDN functions with an open, fully configurable framework would enable joint optimisation of routing, key-pool management and physical reconfiguration, and allow the cross-layer information demonstrated here to be exploited more systematically.
Finally, integrating post-quantum cryptography, particularly for bootstrap authentication and control-plane security, could provide a hybrid security architecture and further demonstrate interoperability between QKD, conventional telecommunications and emerging post-quantum technologies, while making the associated security assumptions explicit.}

\section*{METHODS}\label{sec:methods}
\textbf{Network implementation and application services.}
Each end-user node was equipped with at least one QKD transmitter, one QKD receiver (Quell-X or- XR QKD systems by QTI) and a KMS module (QKME by QTI).
End-users in UC\textsubscript{1} also carry 
Layer~1 encryptors (FSP 3000R7 9TCE-PCN-10GU+AES10G by Adtran), while N2 and N4 are equipped with Layer~3, non-optical, encryptors (TelsyMusaX by Telsy).

All quantum signals are carried by a single fibre while the QKD synchronisation, bidirectional KMS traffic and application services are wavelength multiplexed the second, classical fibre.
At the intermediate ODF sites, spectral or directional separation was applied, while selected KMS signals were optoelectronically regenerated and switched before onward transmission.
The latter enabled the logical mesh KMS connectivity shown in Fig.~1c independently of the direct quantum-link topology.
The Layer~1 encryption service signals were deployed over the classical fibre following the QKD ring service topology, with bidirectional optical paths providing full-duplex 10~Gbit~s$^{-1}$ connectivity between the corresponding nodes.

The QKDN was integrated over existing underground single-mode fibre provided through the Cyprus Telecommunications Authority metropolitan fibre network.
Fibre length and attenuation were characterised before deployment using optical time-domain reflectometry at $1550~\mathrm{nm}$.
The exact wavelength allocation, infrastructure characterisation and full description of the optical implementation are provided in the Supplementary Information.

The Layer~1 and Layer~3 encryption services remained active during network operation and continuously consumed QKD-derived key material from their corresponding link-local key pools.
The Layer~1 encryptors consumed $256~\mathrm{bit~min^{-1}}$ per protected connection to refresh the AES session keys securing the bidirectional optical tunnels.
These encrypted tunnels also carried the monitoring traffic of UC\textsubscript{1}.
In UC\textsubscript{2}, the Layer~3 encryption service between N2 and N4 consumed QKD-derived key material at $128~\mathrm{bit~min^{-1}}$.
The one-time-pad application instead requested end-to-end key material on demand, but its application-level performance was beyond the scope of the present work.

\textbf{Reconfiguration control and latency analysis.}
Physical reconfiguration was confined to the QKD layer and coordinated the quantum and corresponding synchronisation paths.
KMS route selection and physical reconfiguration were implemented as separate control processes.
The reconfiguration controller operated coordinated 2x1 optical switches located at ODF1, selecting either the N2--N4 or N2--N1 configuration.
The deployed rule-based policy evaluated, in priority order, prolonged absence of secret-key generation, stored-key level and the balance between observed key-pool depletion and estimated key production.

During days 35--37 of the monitoring period, the two switch configurations were alternated at a period of 2~h.
For each switching event, the recovery time was measured from the switch-command timestamp to the first subsequent generation of fresh secret-key material by the newly selected QKD link.
Static reference intervals from the same links during periods without optical reconfiguration were analysed using the same fresh-key criterion.

The reported reconfiguration overhead was obtained by comparing the switching and static recovery times and therefore includes synchronisation recovery, QKD-system calibration and key-generation start-up rather than the actuation time of the optical switch alone.

\textbf{End-to-end key-delivery experiment and route reconstruction.}
For each target-volume request interval, completion was defined as the delivery of the full assigned number of 256-bit keys before the next source--destination interval began.
The completion time, $T_j$, was measured from the first key request of interval $j$ to the delivery of its final requested key.
The effective end-to-end key-delivery rate was calculated as
\begin{equation}
R_{\mathrm{E2E},j}
=
\frac{256N_j}{T_j},
\end{equation}
where $N_j$ is the number of 256-bit keys delivered during interval $j$.
Delivery of an end-to-end key consumes stored key material from every physical QKD-link key pool traversed by its trusted-relay path.
Because an explicit per-request route log was not available, physical-link involvement was reconstructed from changes in the link-local key pools, constrained by the known physical QKD topology.
Additional link-pool changes not associated with the reconstructed route were not attributed to the end-to-end request.
Such additional changes could arise from concurrent key consumption by the continuously active Layer~1 and Layer~3 services.
Deliveries for which no compatible route could be assigned were excluded from route-dependent statistics; 6713 of 6789 delivered keys (98.9\%) were route resolved.

For request interval $j$, the involvement of physical QKD link $k$ was defined as 
\begin{equation}
I_{j,k}
=
100\frac{N_{j,k}}{N_{j,\mathrm{res}}},
\end{equation}
where $N_{j,k}$ is the number of route-resolved delivered keys whose reconstructed path traversed physical link $k$, and $N_{j,\mathrm{res}}$ is the total number of route-resolved delivered keys in interval $j$.
Because a single end-to-end delivery may traverse multiple physical links, the involvement percentages within a request need not sum to $100\%$.
The average number of physical QKD hops per route-resolved delivered key was calculated for each request interval from the reconstructed routes.

\textbf{QKD protocol and performance model.}
The deployed QKD systems implement a time-bin-encoded variation of the three-state, one-decoy protocol~\cite{LoMaChen_2005,Ma_2005} at a pulse repetition rate, $f_{rep}$, of $600~\mathrm{MHz}$.
The finite-key calculation followed the composable one-decoy treatment of Refs.~\cite{Rusca_2018,Tomamichel_2012,Lim_2014,Rusca2018ThreeState,Boaron_2018}.
The secret key length, $\ell$, is therefore bounded by
\begin{equation}
\begin{gathered}
\ell \leq
s_{0}^{\mathrm{L}}
+
s_{1}^{\mathrm{L}}
\left[
1-h\!\left(\phi_{\mathrm{1}}^{\mathrm{U}}\right)
\right]
-
f_{\mathrm{EC}}n_{\mathrm{X}}h\left(E_{\mathrm{X}}\right)
- \Delta_{sec},
\\
\Delta_{sec} = 
6\log_{2}\!\left(\frac{19}{\epsilon_{\mathrm{sec}}}\right)
+ 
\log_{2}\!\left(\frac{2}{\epsilon_{\mathrm{cor}}}\right).
\end{gathered}
\label{eq:finite_key_length}
\end{equation}
Here $s_{\mathrm{0}}^{\mathrm{L}}$ and $s_{\mathrm{1}}^{\mathrm{L}}$ are lower bounds on the vacuum and single-photon detection events, respectively, $\phi_{\mathrm{1}}^{\mathrm{U}}$ is an upper bound on the corresponding single-photon phase-error rate and $h$ the binary entropy.
The parameter $f_{\mathrm{EC}}$ is the error-correction factor, here set to 1.14; $n_{\mathrm{X}}$ and $E_{\mathrm{X}}$ are, respectively, the number of detection events and the QBER in the key-generation basis $X$.
The secrecy and correctness parameters were set to $\epsilon_{\mathrm{sec}}=10^{-9}$ and $\epsilon_{\mathrm{cor}}=10^{-15}$, respectively.
Residual correlations between the optical phases of successive pulses were accounted for using the vendor-specified upper bound $p_{\mathrm{phc}} = 0.02$~\cite{Francesconi_2024}, applying the correction 
\begin{equation}
\ell_{\mathrm{final}}
\leq
\left(1-p_{\mathrm{phc}}\right)\ell.
\end{equation}

The finite-key calculation used privacy-amplification blocks containing $n_{\mathrm X}=10^6$ detections in the key-generation basis.
Equal preparation probabilities were assumed for the key-generation and monitoring bases and for the signal and decoy intensities. 
The decoy intensity was fixed 3~dB below the signal intensity, while the signal intensity was optimised at each channel loss to maximise the secure key rate.
The measured total quantum channel loss of each deployed connection was used as the channel input.
A vendor-specified implementation factor of $\eta_{\mathrm{impl}}=0.30$ was applied on the SKR to all links.
Further information on the numerical model and the parameters used are given in the Supplementary Information.

\textbf{Long-term monitoring and data analysis.}
Network performance was monitored continuously for 73 days.
For each active QKD link, the SKR, QBER and generated secret key material were logged from the QKD system and the SDN.
Time-resolved traces were constructed from the logged SKR and QBER values, and cumulative generated-key curves were obtained by summing the generated key material over the monitoring period.

For each link, the mean SKR, mean QBER, and corresponding standard deviations were calculated over the monitoring period using the available logged samples.
SKR histograms were generated from the same time-series data to characterise the distribution of operating points for each link.
The histograms were then fitted using a weighted mixture of two reflected log-normal distributions, with the fitting parameters estimated by bounded non-linear least-squares optimisation.

For permanently provisioned links, an outage was defined as a contiguous interval longer than 1~h during which the link was expected to operate but generated no secret key material.
Availability was calculated as the fraction of scheduled operating time outside these outage intervals.
For the N2--N4 and N2--N1 configurations, periods during which the alternative switch state was intentionally selected were excluded from the active-state availability denominator.

\mm{\textbf{Use of generative AI.}
Generative AI (ChatGPT, OpenAI) was used during manuscript preparation to assist with language, structure and drafting.
All scientific content has been reviewed and verified by the authors.}

\begin{acknowledgments}
This work was co-funded by the European Commission and the Cyprus Deputy Ministry of Research, Innovation and Digital Policy under Grant Agreement No. 101091655.
The funders played no role in study design, data collection, analysis and interpretation of data, or the writing of this manuscript.
\end{acknowledgments}

\section*{Author contributions}
MM designed the network.
AS, MM, SY, KKatzis and EP implemented the network.
AS, MM, performed the experiments, analysed the data and provided the simulations.
SM developed the telemetry software. 
KKalli and MM guided the work.
KKalli coordinated the project.
MM wrote the manuscript, with contributions from all authors.
\medskip

\section*{Competing interests}
All authors declare no competing interests.

\section*{Data availability}
The datasets generated and/or analysed during the current study are not publicly available but are available from the corresponding authors on reasonable request.

\medskip
\section*{Code availability}
The underlying code for this study is not publicly available but may be made available to qualified researchers on reasonable request from the corresponding authors.

\medskip
\section*{Additional information}
\textbf{Correspondence} and requests for materials should be addressed to Mariella Minder and Kyriacos Kalli.

\bibliography{7nodeBiblio}

@inproceedings{Bennett1984,
  author       = {Bennett, Charles H. and Brassard, Gilles},
  title        = {Quantum cryptography: Public key distribution and coin tossing},
  booktitle    = {Proceedings of the IEEE International Conference on Computers, Systems and Signal Processing},
  pages        = {175--179},
  year         = {1984},
  address      = {Bangalore, India}
}

@article{Bogdanski2009,
  author       = {Bogdanski, Jan and Rafiei, Nima and Bourennane, Mohamed},
  title        = {Multiuser Quantum Key Distribution over Telecom Fiber Networks},
  journal      = {Optics Communications},
  volume       = {282},
  number       = {2},
  pages        = {258--262},
  year         = {2009},
  doi          = {10.1016/j.optcom.2008.10.030}
}

@inproceedings{Braun2021OpenQKD,
  author       = {Braun, Ralf-Peter and Geitz, Marc},
  title        = {The {OpenQKD} Testbed in Berlin},
  booktitle    = {Asia Communications and Photonics Conference 2021},
  series       = {Technical Digest Series},
  year         = {2021},
  publisher    = {Optica Publishing Group},
  note         = {Paper M4C.2},
  doi          = {10.1364/ACPC.2021.M4C.2}
}

@article{Chapuran2009,
  author       = {Chapuran, T. E. and Toliver, P. and Peters, N. A. and Jackel, J. and Goodman, M. S. and Runser, R. J. and McNown, S. R. and Dallmann, N. and Hughes, R. J. and McCabe, K. P. and Nordholt, J. E. and Peterson, C. G. and Tyagi, K. T. and Mercer, L. and Dardy, H.},
  title        = {Optical Networking for Quantum Key Distribution and Quantum Communications},
  journal      = {New Journal of Physics},
  volume       = {11},
  pages        = {105001},
  year         = {2009},
  doi          = {10.1088/1367-2630/11/10/105001}
}

@article{Chen2021Hefei,
  author       = {Chen, Teng-Yun and Jiang, Xiao and Tang, Shi-Biao and Zhou, Lei and Yuan, Xiao and Zhou, Hongyi and Wang, Jian and Liu, Yang and Chen, Luo-Kan and Liu, Wei-Yue and Zhang, Hong-Fei and Cui, Ke and Liang, Hao and Li, Xiao-Gang and Mao, Yingqiu and Wang, Liu-Jun and Feng, Si-Bo and Chen, Qing and Zhang, Qiang and Li, Li and Liu, Nai-Le and Peng, Cheng-Zhi and Ma, Xiongfeng and Zhao, Yong and Pan, Jian-Wei},
  title        = {Implementation of a 46-node Quantum Metropolitan Area Network},
  journal      = {npj Quantum Information},
  volume       = {7},
  number       = {1},
  pages        = {134},
  year         = {2021},
  doi          = {10.1038/s41534-021-00474-3}
}

@article{Chen_2021,
  author       = {Yu Ao Chen and Qiang Zhang and Teng Yun Chen and Wen Qi Cai and Sheng Kai Liao and Jun Zhang and Kai Chen and Juan Yin and Ji Gang Ren and Zhu Chen and Sheng Long Han and Qing Yu and Ken Liang and Fei Zhou and Xiao Yuan and Mei Sheng Zhao and Tian Yin Wang and Xiao Jiang and Liang Zhang and Wei Yue Liu and Yang Li and Qi Shen and Yuan Cao and Chao Yang Lu and Rong Shu and Jian Yu Wang and Li Li and Nai Le Liu and Feihu Xu and Xiang Bin Wang and Cheng Zhi Peng and Jian Wei Pan},
  title        = {An integrated space-to-ground quantum communication network over 4,600 kilometres},
  journal      = {Nature},
  volume       = {589},
  number       = {7841},
  pages        = {214--219},
  year         = {2021},
  doi          = {10.1038/s41586-020-03093-8}
}

@article{Chen_2025,
  author       = {Chen, Hao-Ze and Li, Ming-Han and Wang, Yu Zhou and Zhao, Zhen-Geng and Ye, Cheng and Li, Fei Long and Chen, Zhu and Han, Sheng-Long and Tang, Bao and Miao, Ya Jun and Qi, Wei},
  title        = {Implementation of carrier-grade quantum communication networks over 10000 km},
  journal      = {npj Quantum Information},
  volume       = {11},
  pages        = {137},
  year         = {2025},
  doi          = {10.1038/s41534-025-01089-8}
}

@inproceedings{Cho2021,
  author       = {Cho, Joo Yeon and Pedreno-Manresa, Jose-Juan and Patri, Sai and Sergeev, Andrew and Elbers, J{\"o}rg-Peter and Griesser, Helmut and White, Catherine and Lord, Andrew},
  title        = {Demonstration of Software-defined Key Management for Quantum Key Distribution Network},
  booktitle    = {Optical Fiber Communication Conference (OFC) 2021},
  year         = {2021},
  publisher    = {Optica Publishing Group},
  note         = {Paper M2B.4},
  doi          = {10.1364/OFC.2021.M2B.4}
}

@article{Dynes2019,
  author       = {J. F. Dynes and A. Wonfor and W. W.-S. Tam and A. W. Sharpe and R. Takahashi and M. Lucamarini and A. Plews and Z. L. Yuan and A. R. Dixon and J. Cho and Y. Tanizawa and J.-P. Elbers and H. Grei{\ss}er and I. H. White and R. V. Penty and A. J. Shields},
  title        = {Cambridge quantum network},
  journal      = {npj Quantum Information},
  volume       = {5},
  number       = {1},
  pages        = {101},
  year         = {2019},
  doi          = {10.1038/s41534-019-0221-4}
}

@inproceedings{Elliot_2005,
  author       = {Elliott, Chip and Colvin, Alexander and Pearson, David and Pikalo, Oleksiy and Schlafer, John and Yeh, Henry},
  title        = {Current status of the {DARPA} Quantum Network},
  booktitle    = {Quantum Information and Computation III},
  series       = {Proceedings of SPIE},
  volume       = {5815},
  pages        = {138--149},
  year         = {2005},
  doi          = {10.1117/12.606489}
}

@article{Eraerds2010,
  author       = {Eraerds, Patrick and Walenta, Nino and Legr{\'e}, Matthieu and Gisin, Nicolas and Zbinden, Hugo},
  title        = {Quantum Key Distribution and 1 Gbit/s Data Encryption over a Single Fibre},
  journal      = {New Journal of Physics},
  volume       = {12},
  number       = {6},
  pages        = {063027},
  year         = {2010},
  doi          = {10.1088/1367-2630/12/6/063027}
}

@article{Francesconi_2024,
  author       = {Francesconi, Saverio and De Lazzari, Claudia and Ribezzo, Domenico and Vagniluca, Ilaria and Biagi, Nicola and Occhipinti, Tommaso and Zavatta, Alessandro and Bacco, Davide},
  title        = {Scalable Implementation of Temporal and Phase Encoding {QKD} with Phase-Randomized States},
  journal      = {Advanced Quantum Technologies},
  volume       = {7},
  number       = {2},
  pages        = {2300224},
  year         = {2024},
  doi          = {10.1002/qute.202300224}
}

@article{Frohlich2013,
  author       = {Fr{\"o}hlich, Bernd and Dynes, James F. and Lucamarini, Marco and Sharpe, Andrew W. and Yuan, Zhiliang and Shields, Andrew J.},
  title        = {A Quantum Access Network},
  journal      = {Nature},
  volume       = {501},
  number       = {7465},
  pages        = {69--72},
  year         = {2013},
  doi          = {10.1038/nature12493}
}

@article{Gisin2002,
  author       = {Gisin, Nicolas and Ribordy, Gr{\'e}goire and Tittel, Wolfgang and Zbinden, Hugo},
  title        = {Quantum cryptography},
  journal      = {Reviews of Modern Physics},
  volume       = {74},
  number       = {1},
  pages        = {145--195},
  year         = {2002},
  doi          = {10.1103/RevModPhys.74.145}
}

@article{Liao2017,
  author       = {Liao, Sheng-Kai and Cai, Wen-Qi and Liu, Wei-Yue and Zhang, Liang and Li, Yang and Ren, Ji-Gang and Yin, Juan and Shen, Qi and Cao, Yuan and Li, Zheng-Ping and Li, Feng-Zhi and Chen, Xia-Wei and Sun, Li-Hua and Jia, Jian-Jun and Wu, Jin-Cai and Jiang, Xiao-Jun and Wang, Jian-Feng and Huang, Yong-Mei and Wang, Qiang and Zhou, Yi-Lin and Deng, Lei and Xi, Tao and Ma, Lu and Hu, Tai and Zhang, Qiang and Chen, Yu-Ao and Liu, Nai-Le and Wang, Xiang-Bin and Zhu, Zhen-Cai and Lu, Chao-Yang and Shu, Rong and Peng, Cheng-Zhi and Wang, Jian-Yu and Pan, Jian-Wei},
  title        = {Satellite-to-ground quantum key distribution},
  journal      = {Nature},
  volume       = {549},
  number       = {7670},
  pages        = {43--47},
  year         = {2017},
  doi          = {10.1038/nature23655}
}

@article{Lim_2014,
  author       = {Lim, Charles Ci Wen and Curty, Marcos and Walenta, Nino and Xu, Feihu and Zbinden, Hugo},
  title        = {Concise Security Bounds for Practical Decoy-State Quantum Key Distribution},
  journal      = {Physical Review A},
  volume       = {89},
  number       = {2},
  pages        = {022307},
  year         = {2014},
  doi          = {10.1103/PhysRevA.89.022307}
}

@article{Lo2014,
  author       = {Lo, Hoi-Kwong and Curty, Marcos and Tamaki, Kiyoshi},
  title        = {Secure Quantum Key Distribution},
  journal      = {Nature Photonics},
  volume       = {8},
  number       = {8},
  pages        = {595--604},
  year         = {2014},
  doi          = {10.1038/nphoton.2014.149}
}

@article{LoMaChen_2005,
  author       = {Lo, Hoi-Kwong and Ma, Xiongfeng and Chen, Kai},
  title        = {Decoy State Quantum Key Distribution},
  journal      = {Physical Review Letters},
  volume       = {94},
  number       = {23},
  pages        = {230504},
  year         = {2005},
  doi          = {10.1103/PhysRevLett.94.230504}
}

@article{Lopez2020,
  author       = {Lopez, Diego R. and Martin, Vicente and Lopez, Victor and de la Iglesia, Fernando and Pastor, Antonio and Brunner, Hans and Aguado, Alejandro and Bettelli, Stefano and Fung, Fred and Hillerkuss, David and Comandar, Lucian and Wang, Dong and Poppe, Andreas and Brito, Juan P. and Salas, Pedro J. and Peev, Momtchil},
  title        = {Demonstration of Software Defined Network Services Utilizing Quantum Key Distribution Fully Integrated with Standard Telecommunication Network},
  journal      = {Quantum Reports},
  volume       = {2},
  number       = {3},
  pages        = {453--458},
  year         = {2020},
  doi          = {10.3390/quantum2030032}
}

@article{Ma2007,
  author       = {Ma, Lijun and Mink, Alan and Xu, Hai and Slattery, Oliver and Tang, Xiao},
  title        = {Experimental Demonstration of an Active Quantum Key Distribution Network with Over Gbps Clock Synchronization},
  journal      = {IEEE Communications Letters},
  volume       = {11},
  number       = {12},
  pages        = {1019--1021},
  year         = {2007},
  doi          = {10.1109/LCOMM.2007.071477}
}

@article{Ma_2005,
  author       = {Ma, Xiongfeng and Qi, Bing and Zhao, Yi and Lo, Hoi-Kwong},
  title        = {Practical Decoy State for Quantum Key Distribution},
  journal      = {Physical Review A},
  volume       = {72},
  number       = {1},
  pages        = {012326},
  year         = {2005},
  doi          = {10.1103/PhysRevA.72.012326}
}

@article{Mao2018,
  author       = {Mao, Yingqiu and Wang, Bi-Xiao and Zhao, Chunxu and Wang, Guangquan and Wang, Ruichun and Wang, Honghai and Zhou, Fei and Nie, Jimin and Chen, Qing and Zhao, Yong and Zhang, Qiang and Zhang, Jun and Chen, Teng-Yun and Pan, Jian-Wei},
  title        = {Integrating Quantum Key Distribution with Classical Communications in Backbone Fiber Network},
  journal      = {Optics Express},
  volume       = {26},
  number       = {5},
  pages        = {6010--6020},
  year         = {2018},
  doi          = {10.1364/OE.26.006010}
}

@article{Martin2024,
  author       = {Martin, V. and Brito, J. P. and Ort{\'i}z, L. and M{\'e}ndez, R. B. and Buruaga, J. S. and Vicente, R. J. and Sebasti{\'a}n-Lombra{\~n}a, A. and Rinc{\'o}n, D. and P{\'e}rez, F. and S{\'a}nchez, C. and Peev, M. and Brunner, H. H. and Fung, F. and Poppe, A. and Fr{\"o}wis, F. and Shields, A. J. and Woodward, R. I. and Griesser, H. and Roehrich, S. and de la Iglesia, F. and Abell{\'a}n, C. and Hentschel, M. and Rivas-Moscoso, J. M. and Pastor-Perales, A. and Folgueira, J. and L{\'o}pez, D.},
  title        = {{MadQCI}: A Heterogeneous and Scalable {SDN-QKD} Network Deployed in Production Facilities},
  journal      = {npj Quantum Information},
  volume       = {10},
  number       = {1},
  pages        = {80},
  year         = {2024},
  doi          = {10.1038/s41534-024-00873-2}
}

@article{Mehic2020Networking,
  author       = {Mehic, Miralem and Niemiec, Marcin and Rass, Stefan and Ma, Jiajun and Peev, Momtchil and Aguado, Alejandro and Martin, Vicente and Schauer, Stefan and Poppe, Andreas and Pacher, Christoph and Voznak, Miroslav},
  title        = {Quantum Key Distribution: A Networking Perspective},
  journal      = {ACM Computing Surveys},
  volume       = {53},
  number       = {5},
  pages        = {96:1--96:41},
  year         = {2020},
  doi          = {10.1145/3402192}
}

@inproceedings{Ou2018,
  author       = {Ou, Yanni and Hugues-Salas, Emilio and Ntavou, Foteini and Wang, Rui and Bi, Yu and Yan, Shuangyi and Kanellos, George T. and Nejabati, Reza and Simeonidou, Dimitra},
  title        = {Field-Trial of Machine Learning-Assisted Quantum Key Distribution ({QKD}) Networking with {SDN}},
  booktitle    = {2018 European Conference on Optical Communication (ECOC)},
  pages        = {1--3},
  year         = {2018},
  publisher    = {IEEE},
  doi          = {10.1109/ECOC.2018.8535497}
}

@article{Patel2012,
  author       = {Patel, K. A. and Dynes, J. F. and Choi, I. and Sharpe, A. W. and Dixon, A. R. and Yuan, Z. L. and Penty, R. V. and Shields, A. J.},
  title        = {Coexistence of High-Bit-Rate Quantum Key Distribution and Data on Optical Fiber},
  journal      = {Physical Review X},
  volume       = {2},
  number       = {4},
  pages        = {041010},
  year         = {2012},
  doi          = {10.1103/PhysRevX.2.041010}
}

@article{Peev_2009,
  author       = {Peev, M and Pacher, C and All{\'e}aume, R and Barreiro, C and Bouda, J and Boxleitner, W and Debuisschert, T and Diamanti, E and Dianati, M and Dynes, J F and Fasel, S and Fossier, S and F{\"u}rst, M and Gautier, J-D and Gay, O and Gisin, N and Grangier, P and Happe, A and Hasani, Y and Hentschel, M and H{\"u}bel, H and Humer, G and L{\"a}nger, T and Legr{\'e}, M and Lieger, R and Lodewyck, J and Lor{\"u}nser, T and L{\"u}tkenhaus, N and Marhold, A and Matyus, T and Maurhart, O and Monat, L and Nauerth, S and Page, J-B and Poppe, A and Querasser, E and Ribordy, G and Robyr, S and Salvail, L and Sharpe, A W and Shields, A J and Stucki, D and Suda, M and Tamas, C and Themel, T and Thew, R T and Thoma, Y and Treiber, A and Trinkler, P and Tualle-Brouri, R and Vannel, F and Walenta, N and Weier, H and Weinfurter, H and Wimberger, I and Yuan, Z L and Zbinden, H and Zeilinger, A},
  title        = {The {SECOQC} quantum key distribution network in Vienna},
  journal      = {New Journal of Physics},
  volume       = {11},
  number       = {7},
  pages        = {075001},
  year         = {2009},
  doi          = {10.1088/1367-2630/11/7/075001}
}

@inproceedings{Poppe2007,
  author       = {Poppe, Andreas and H{\"u}bel, Hannes and Karinou, Fotini and Blauensteiner, Bibiane and Schrenk, Bernhard and Lor{\"u}nser, Thomas and Meyenburg, Markus and Querasser, Erich and Zeilinger, Anton},
  title        = {Quantum Key Distribution over {WDM}s and Optical Switches to Combine the Quantum Channel with Synchronization Channels},
  booktitle    = {33rd European Conference and Exhibition of Optical Communication (ECOC 2007)},
  pages        = {1--2},
  year         = {2007},
  publisher    = {VDE},
  address      = {Berlin},
  doi          = {10.1049/ic:20070343}
}

@article{Rusca_2018,
  author       = {Rusca, Davide and Boaron, Alberto and Gr{\"u}nenfelder, Fadri and Martin, Anthony and Zbinden, Hugo},
  title        = {Finite-Key Analysis for the 1-Decoy State {QKD} Protocol},
  journal      = {Applied Physics Letters},
  volume       = {112},
  number       = {17},
  pages        = {171104},
  year         = {2018},
  doi          = {10.1063/1.5023340}
}

@article{Sasaki_2011,
  author       = {M. Sasaki and M. Fujiwara and H. Ishizuka and W. Klaus and K. Wakui and M. Takeoka and S. Miki and T. Yamashita and Z. Wang and A. Tanaka and K. Yoshino and Y. Nambu and S. Takahashi and A. Tajima and A. Tomita and T. Domeki and T. Hasegawa and Y. Sakai and H. Kobayashi and T. Asai and K. Shimizu and T. Tokura and T. Tsurumaru and M. Matsui and T. Honjo and K. Tamaki and H. Takesue and Y. Tokura and J. F. Dynes and A. R. Dixon and A. W. Sharpe and Z. L. Yuan and A. J. Shields and S. Uchikoga and M. Legr\'{e} and S. Robyr and P. Trinkler and L. Monat and J.-B. Page and G. Ribordy and A. Poppe and A. Allacher and O. Maurhart and T. L\"{a}nger and M. Peev and A. Zeilinger},
  title        = {Field test of quantum key distribution in the Tokyo {QKD} Network},
  journal      = {Optics Express},
  volume       = {19},
  number       = {11},
  pages        = {10387--10409},
  year         = {2011},
  doi          = {10.1364/OE.19.010387}
}

@article{Scarani2009,
  author       = {Scarani, Valerio and Bechmann-Pasquinucci, Helle and Cerf, Nicolas J. and Du{\v{s}}ek, Miloslav and L{\"u}tkenhaus, Norbert and Peev, Momtchil},
  title        = {The security of practical quantum key distribution},
  journal      = {Reviews of Modern Physics},
  volume       = {81},
  number       = {3},
  pages        = {1301--1350},
  year         = {2009},
  doi          = {10.1103/RevModPhys.81.1301}
}

@article{Stucki_2011,
  author       = {Stucki, D and Legr{\'e}, M and Buntschu, F and Clausen, B and Felber, N and Gisin, N and Henzen, L and Junod, P and Litzistorf, G and Monbaron, P and Monat, L and Page, J-B and Perroud, D and Ribordy, G and Rochas, A and Robyr, S and Tavares, J and Thew, R and Trinkler, P and Ventura, S and Voirol, R and Walenta, N and Zbinden, H},
  title        = {Long-term performance of the SwissQuantum quantum key distribution network in a field environment},
  journal      = {New Journal of Physics},
  volume       = {13},
  number       = {12},
  pages        = {123001},
  year         = {2011},
  doi          = {10.1088/1367-2630/13/12/123001}
}

@inproceedings{Tessinari2019,
  author       = {Tessinari, Rodrigo S. and Bravalheri, Anderson and Hugues-Salas, Emilio and Collins, Richard and Aktas, Djeylan and Guimaraes, Rafael S. and Alia, Obada and Rarity, John and Kanellos, George T. and Nejabati, Reza and Simeonidou, Dimitra},
  title        = {Field Trial of Dynamic {DV-QKD} Networking in the {SDN}-Controlled Fully-Meshed Optical Metro Network of the Bristol City {5GUK} Test Network},
  booktitle    = {45th European Conference on Optical Communication (ECOC 2019)},
  year         = {2019},
  publisher    = {Institution of Engineering and Technology},
  doi          = {10.1049/cp.2019.1033}
}

@article{Tomamichel_2012,
  author       = {Tomamichel, Marco and Lim, Charles Ci Wen and Gisin, Nicolas and Renner, Renato},
  title        = {Tight Finite-Key Analysis for Quantum Cryptography},
  journal      = {Nature Communications},
  volume       = {3},
  pages        = {634},
  year         = {2012},
  doi          = {10.1038/ncomms1631}
}

@article{Wang2010Wavelength,
  author       = {Wang, Shuang and Chen, Wei and Yin, Zhen-Qiang and Zhang, Yang and Zhang, Tao and Li, Hong-Wei and Xu, Fang-Xing and Zhou, Zheng and Yang, Yang and Huang, Da-Jun and Zhang, Li-Jun and Li, Fang-Yi and Liu, Dong and Wang, Yong-Gang and Guo, Guang-Can and Han, Zheng-Fu},
  title        = {Field Test of Wavelength-Saving Quantum Key Distribution Network},
  journal      = {Optics Letters},
  volume       = {35},
  number       = {14},
  pages        = {2454--2456},
  year         = {2010},
  doi          = {10.1364/OL.35.002454}
}

@article{Xu2009,
  author       = {Xu, Fangxing and Chen, Wei and Wang, Shuang and Yin, Zhenqiang and Zhang, Yang and Liu, Yun and Zhou, Zheng and Zhao, Yibo and Li, Hongwei and Liu, Dong and Han, Zhengfu and Guo, Guangcan},
  title        = {Field Experiment on a Robust Hierarchical Metropolitan Quantum Cryptography Network},
  journal      = {Chinese Science Bulletin},
  volume       = {54},
  number       = {17},
  pages        = {2991--2997},
  year         = {2009},
  doi          = {10.1007/s11434-009-0526-3}
}

@inproceedings{Minder2025SPIE,
  author    = {Minder, Mariella and Siakolas, Andreas and Yerolatsitis, Stephanos and Katzis, Konstantinos and Kalli, Kyriacos},
  title     = {Multi-node Quantum Key Distribution Network Using Existing Underground Optical Fibre Infrastructure},
  booktitle = {Eighth International Workshop on Specialty Optical Fibers and Their Applications},
  series    = {Proceedings of SPIE},
  volume    = {13522},
  pages     = {1352219},
  year      = {2025},
  doi       = {10.1117/12.3066715}
}

@article{Dijkstra1959,
  author  = {Dijkstra, E. W.},
  title   = {A Note on Two Problems in Connexion with Graphs},
  journal = {Numerische Mathematik},
  volume  = {1},
  pages   = {269--271},
  year    = {1959},
  doi     = {10.1007/BF01386390}
}

@misc{DeLazzari2026Milan,
  author       = {De Lazzari, Claudia and Biagi, Nicola and Giani, Damiano
                  and Russo, Marco and Chirici, Fernando and Stocco, Francesco
                  and Francesconi, Saverio and Ferranti, Giacomo
                  and Soureal, Alessandro and Sanguineti, Antonella
                  and Montrucchio, Bartolomeo and Laurenzi, Christian
                  and Testa, Oliviero and Morgari, Guglielmo
                  and Manzalini, Antonio and Occhipinti, Tommaso
                  and Zavatta, Alessandro and Bacco, Davide},
  title        = {Deploying and validating a metropolitan {QKD} secure network:
                  architecture and field performance},
  year         = {2026},
  eprint       = {2607.11727},
  archivePrefix = {arXiv},
  primaryClass = {quant-ph}
}

@article{AmiesKing2023,
  author  = {Amies-King, Ben and Schatz, Karolina P. and Duan, Haofan and
             Biswas, Ayan and Bailey, Jack and Felvinti, Adrian and
             Winward, Jaimes and Dixon, Mike and Minder, Mariella and
             Kumar, Rupesh and Albosh, Sophie and Lucamarini, Marco},
  title   = {Quantum Communications Feasibility Tests over a UK-Ireland
             224 km Undersea Link},
  journal = {Entropy},
  volume  = {25},
  number  = {12},
  pages   = {1572},
  year    = {2023},
  doi     = {10.3390/e25121572}
}

@article{Rusca2018ThreeState,
   author = {Davide Rusca and Alberto Boaron and Marcos Curty and Anthony Martin and Hugo Zbinden},
   doi = {10.1103/PhysRevA.98.052336},
   issn = {24699934},
   issue = {5},
   journal = {Physical Review A},
   month = {11},
   publisher = {American Physical Society},
   title = {Security proof for a simplified Bennett-Brassard 1984 quantum-key-distribution protocol},
   volume = {98},
   year = {2018}
}

@article{Boaron_2018,
   title={Simple 2.5~GHz time-bin quantum key distribution},
   volume={112},
   ISSN={1077-3118},
   url={http://dx.doi.org/10.1063/1.5027030},
   DOI={10.1063/1.5027030},
   number={17},
   journal={Applied Physics Letters},
   publisher={AIP Publishing},
   author={Boaron, Alberto and Korzh, Boris and Houlmann, Raphael and Boso, Gianluca and Rusca, Davide and Gray, Stuart and Li, Ming-Jun and Nolan, Daniel and Martin, Anthony and Zbinden, Hugo},
   year={2018},
   month=Apr }

\clearpage
\pagestyle{plain}

\onecolumngrid
\raggedbottom

\begin{center}

{\Large\bfseries
Supplementary Information
\par}

\vspace{1.2em}

{\Large
A Reconfigurable Multilayer Quantum Key Distribution Network\\
over Existing Metropolitan Fibre
\par}

\vspace{1.5em}

\begin{minipage}{0.94\textwidth}
\centering
Mariella Minder,
Andreas Siakolas,
Elizabeth Pasatembou,
Stylianos Mavrikos,
Stephanos Yerolatsitis,
Konstantinos Katzis
\& Kyriacos Kalli

\end{minipage}

\vspace{1.2em}

\begin{minipage}{0.90\textwidth}
\centering
\small
PhOSLab, Department of Electrical Engineering, Computer Engineering and Informatics, Cyprus University of Technology, Limassol 3036, Cyprus\\
\end{minipage}

\end{center}

\vspace{1.5em}



\setcounter{section}{0}
\setcounter{subsection}{0}
\setcounter{figure}{0}
\setcounter{table}{0}
\setcounter{equation}{0}

\renewcommand{\thesection}{S\arabic{section}}
\renewcommand{\thesubsection}{S\arabic{section}.\arabic{subsection}}
\renewcommand{\thefigure}{S\arabic{figure}}
\renewcommand{\thetable}{S\arabic{table}}
\renewcommand{\theequation}{S\arabic{equation}}

\section{Supplementary Note 1: Detailed optical implementation}
\subsection{Network resources and wavelength allocation}
The deployed network comprises seven end-user nodes, N1--N7, connected through four intermediate optical distribution frame sites, ODF1--ODF4.
Each node is equipped with at least one QKD transmitter, $Q_{A,n}$, and one QKD receiver, $Q_{B,n}$, where $n$ identifies the corresponding QKD system.
Each QKD system uses a quantum-signal wavelength, $\lambda_{q,n}$, and a synchronisation wavelength, $\lambda_{s,n}$.
A single key management system (KMS) is also installed at each node and uses the bidirectional optical wavelengths $\lambda_{\mathrm{Tx}}$ and $\lambda_{\mathrm{Rx}}$.
In all nodes, the KMS optical interfaces utilise $\lambda_{\mathrm{Tx}}=\mathrm{ch}\,33, \lambda_{\mathrm{Rx}}=\mathrm{ch}\,32$.
Exact resource and wavelength allocation for the quantum layers is given in Supplementary Table~\ref{tab:sup:allocation}, which reports the nominal dense wavelength division multiplexing (DWDM) channel assigned to each carrier and, for the revised use case 1 (UC\textsubscript{1}) implementation, the four quantum carriers' central wavelengths within  channel~40.

Layer~1 encryptors are also deployed as part of the application layer of UC\textsubscript{1}, at N1, N5, N6 and N7.
They create 10~Gbit~s$^{-1}$ bidirectional secure tunnels between the nodes over the same ring topology of the quantum layer.
Layer~1 wavelength channel allocation is provided in Supplementary Table~\ref{tab:sup:allocation_encryptors}.
Layer~3 encryptors, deployed at N2 and N4 only act at the IP sec level and therefore have no optical traffic.
Throughout the Supplementary Information, channel numbers refer to the 100~GHz ITU–T DWDM grid in the C-band.
The on-demand one-time-pad application similarly operates over the existing classical network and requests key material through the KMS application interface without generating optical traffic.

\begin{table}[H]
    \centering
    \caption{\textbf{Quantum layer wavelength allocation.}
    Wavelengths are given as channels of the 100~GHz ITU–T DWDM grid in the C-band.
    For UC\textsubscript{1}, the central wavelength of each link is also given, as the carriers are spectrally offset within the nominal DWDM channel.
    The alternative receivers $Q_{B,5}$ and $Q_{B,8}$ share transmitter $Q_{A,5}$ and are selected by the optical switch located at ODF1.}
    \label{tab:sup:allocation}
    \renewcommand{\arraystretch}{1.15}
    \setlength{\tabcolsep}{6pt}
    \begin{tabular}{cclcc}
        \toprule
        \textbf{Link} &
        \textbf{QKD transmitter} &
        \textbf{QKD receiver} &
        \textbf{Synchronisation signal} &
        \textbf{Quantum signal} \\
        \midrule
        N1--N7 &
        $Q_{A,1}$ at N1 &
        $Q_{B,1}$ at N7 &
        $\lambda_{s,1}=\mathrm{ch}\,34$ &
        $\lambda_{q,1}=\mathrm{ch}\,40$ $(1545.191$ nm) \\[2pt]
        N7--N6 &
        $Q_{A,2}$ at N7 &
        $Q_{B,2}$ at N6 &
        $\lambda_{s,2}=\mathrm{ch}\,20$ &
        $\lambda_{q,2}=\mathrm{ch}\,40$ $(1544.991$ nm)\\[2pt]
        N6--N5 &
        $Q_{A,3}$ at N6 &
        $Q_{B,3}$ at N5 &
        $\lambda_{s,3}=\mathrm{ch}\,22$ &
        $\lambda_{q,3}=\mathrm{ch}\,40$ $(1545.679$ nm)\\[2pt]
        N5--N1 &
        $Q_{A,4}$ at N5 &
        $Q_{B,4}$ at N1 &
        $\lambda_{s,4}=\mathrm{ch}\,24$ &
        $\lambda_{q,4}=\mathrm{ch}\,40$ $(1545.432$ nm)\\[2pt]
        N2--N4 &
        {$Q_{A,5}$ at N2} &
        $Q_{B,5}$ at N4 &
        {$\lambda_{s,5}=\mathrm{ch}\,35$} &
        {$\lambda_{q,5}=\mathrm{ch}\,34$} \\[2pt]
        N2--N1 &
        {$Q_{A,5}$ at N2}&
        $Q_{B,8}$ at N1 &
        {$\lambda_{s,5}=\mathrm{ch}\,35$} &
        {$\lambda_{q,5}=\mathrm{ch}\,34$} \\[2pt]
        N4--N3 &
        $Q_{A,6}$ at N4 &
        $Q_{B,6}$ at N3 &
        $\lambda_{s,6}=\mathrm{ch}\,23$ &
        $\lambda_{q,6}=\mathrm{ch}\,36$ \\[2pt]
        N3--N2 &
        $Q_{A,7}$ at N3 &
        $Q_{B,7}$ at N2 &
        $\lambda_{s,7}=\mathrm{ch}\,21$ &
        $\lambda_{q,7}=\mathrm{ch}\,22$ \\[2pt]
        \bottomrule
    \end{tabular}
\end{table}

\begin{table}[H]
    \centering
    \caption{\textbf{Layer~1 encryptors wavelength allocation.}
    The signals are bidirectional forming a ring architecture between nodes N1-N7, N7-N6, N6-N5, N5-N1.}
    \label{tab:sup:allocation_encryptors}
    \renewcommand{\arraystretch}{1.15}
    \setlength{\tabcolsep}{6pt}
    \begin{tabular}{cclcc}
        \toprule
        \textbf{Link} &
        \textbf{Transmitter signal}\\
        \midrule
        $N1 \rightarrow N7$ &
        $\lambda_{L_{1,7}}=\mathrm{ch}\,30$ \\[2pt]
        $N7 \rightarrow N1$ &
        $\lambda_{L_{7,1}}=\mathrm{ch}\,31$\\[2pt]
        \midrule
        $N7 \rightarrow N6$ &
        $\lambda_{L_{7,6}}=\mathrm{ch}\,28$\\[2pt]
        $N6 \rightarrow N7$ &
        $\lambda_{L_{6,7}}=\mathrm{ch}\,29$\\[2pt]
        \midrule
        $N6 \rightarrow N5$ &
        $\lambda_{L_{6,5}}=\mathrm{ch}\,30$\\[2pt]
        $N5 \rightarrow N6$ &
        $\lambda_{L_{5,6}}=\mathrm{ch}\,31$\\[2pt]
        \midrule
        $N5 \rightarrow N1$ &
        $\lambda_{L_{5,1}}=\mathrm{ch}\,28$\\[2pt]
        $N1 \rightarrow N5$ &
        $\lambda_{L_{1,5}}=\mathrm{ch}\,29$\\[2pt]
        
        \bottomrule
    \end{tabular}
\end{table}

\subsection{Signal-plane routing and reconfiguration}
The network is comprised of three distinct layers: quantum, key management and application, which are carried between locations on two separate single-mode fibres.
Quantum signals are dense wavelength division multiplexed over a dedicated dark fibre (quantum channel), while QKD synchronisation, key management and application traffic are multiplexed over the second fibre (classical channel).
Functionally, the QKD layer comprises both the quantum signal and its corresponding synchronisation signal, although these two signals are carried on separate physical fibres in the deployed architecture.
The port-level optical architecture of both the quantum and classical channels across the entire, seven-node network is given in Supplementary Fig~\ref{fig:sup:portLevel} and is summarised below.

Quantum signals are treated differently in UC\textsubscript{1} and UC\textsubscript{2}.
In the quantum channel of UC\textsubscript{1}, all quantum signals share the same DWDM channel.
Optical circulators installed at the end-user nodes and intermediate ODF sites establish a unidirectional quantum ring over the available branched fibre plant as N1-N7-N6-N5-N1.
However, finite circulator isolation gives rise to leakage between co-located transmitters and receivers, and the transmitter frequencies were therefore adjusted by temperature tuning and separated into approximately $10~\mathrm{GHz}$ spectral sub-bands of the nominal DWDM channel.
Narrowband $10~\mathrm{GHz}$ filters were placed at the receivers to pass the spectral sub-band of the intended remote transmitter while suppressing leakage from the transmitter located at the same node.
In UC\textsubscript{2}, the quantum signals between N2,N3 and N4 are carried on distinct DWDM channels.
Here direction is again provided by utilising circulators located at the relevant nodes and ODFs.

Even though all bright signals across the three layers share the same deployed fibres, they follow different logical routing arrangements.
The synchronisation signal associated with each QKD pair follows the same end-to-end route as its corresponding quantum signal.
However, the synchronisation wavelengths are multiplexed onto the classical fibre and are added, dropped or redirected at the intermediate ODF sites.
For the reconfigurable $Q_{A,5}$ transmitter at N2, the synchronisation signal is demultiplexed and optically switched together with the quantum signal.
The selected switch state therefore establishes both the quantum connection and the corresponding synchronisation connection towards either $Q_{B,5}$ at N4 or $Q_{B,8}$ at N1.
Instead, the independent, bidirectional signals of the KMSs, are optoelectronically regenerated and routed at the intermediate ODF sites to provide logical mesh connectivity as described in the main text.
Application-layer signals associated with the Layer~1 encryptors in UC\textsubscript{1} are also transported over the classical fibre.
In this case, the encryptors form a ring following the UC\textsubscript{1} node sequence N1--N7--N6--N5--N1, similar to the quantum-layer topology but carrying traffic in both directions.
At ODF4, four DWDM filters separate and route the application wavelengths associated with the ODF2-, N5-, N6- and N7-facing classical-fibre segments.

Accommodating the optical switching of the quantum and synchronisation signals between the transmitter A5 at N2 and the receivers B5 at N4 and B8 at N1, required the use of a duplex 2x1 optical switch (OS) located at ODF1, acting on both the quantum and the classical fibres.
In one state, the quantum output of A5 and its synchronisation signal are routed through ODF2 and ODF3 towards B5 at N4.
In the alternative state, both signals are redirected towards B8 at N1.
The two configurations are mutually exclusive because they share A5 and the ODF1 switch resources.
The permanent UC\textsubscript{2} links are unaffected by this reconfiguration, and the logical KMS mesh remains available in both switch states.
Following a switch operation, the newly selected QKD pair must recover synchronisation, perform calibration and accumulate sufficient detection statistics before generating fresh secret-key material.

Wavelength-selective components were also necessary at N1 and ODF1 in order to separate the UC\textsubscript{1} quantum signals in nominal channel~40 from the switched N2--N1 quantum signal in channel 34, allowing the two signal groups to share the N1–ODF1 quantum-fibre segment.
Within UC\textsubscript{2}, the N3--N2 link required additional routing because the quantum signal from $A_7$ at N3 propagates towards $B_7$ at N2, opposite to the N2-originating signal from $A_5$ over the fibre segments shared by the two links. Circulators and wavelength-selective add/drop components therefore separate and direct the channel~22 and channel~34 quantum signals at the intermediate ODF sites while allowing them to share the same physical fibre.
Wavelength-selective components at ODF1, ODF2 and ODF3 route the UC\textsubscript{2} signals between N2, N3 and N4. ODF3 provides the principal UC\textsubscript{2} aggregation point, connecting the ODF2-facing quantum fibre to the N3- and N4-facing fibres. Quantum routing at all intermediate ODF sites is passive, with no detection or regeneration of the quantum signals.

\subsection{Optical losses}

The deployed fibres carrying the QKD signals were characterised using optical time-domain reflectometry at $1550~\mathrm{nm}$.
The deployed-fibre loss, $\mathcal{L}_{\mathrm{fib}}$, includes the underground fibre, connectors and patch panels along the selected route.
The total quantum channel loss, $\mathcal{L}$, additionally includes the insertion losses of the optical circulators, narrowband filters, wavelength-selective components and, for the reconfigurable links, the optical switch.
The total quantum channel loss was obtained from the single-photon-level measurements performed by the QKD systems to optimise their protocol implementation parameters.
The additional passive-component loss is therefore
$\mathcal{L}_{\mathrm{comp}}
=
\mathcal{L}
-
\mathcal{L}_{\mathrm{fib}}.$
$\mathcal{L}_{\mathrm{comp}}$ should consequently be interpreted as the aggregate additional loss introduced by the passive network components. 
The total fibre lengths and corresponding network attenuation values are summarised in Supplementary Table~\ref{tab:supp_losses}.
The measured total quantum channel losses, $\mathcal{L}$, were used as the channel loss
inputs to the QKD performance model presented in the main text.

\begin{table}[H]
    \centering
    \caption{\textbf{Quantum channel loss characterisation.}
    While the deployed fibre loss, $\mathcal{L}_{\mathrm{fib}}$ includes only the losses from the inherited fibre infrastructure, $\mathcal{L}$ provides the total loss experienced by the quantum signals due to the use of additional optical components.
    $\mathcal{L}_{\mathrm{comp}}$ is the additional loss introduced by the latter, and is calculated by subtracting the two measured quantities, $\mathcal{L}_{\mathrm{fib}}$ from $\mathcal{L}$.}
    \label{tab:supp_losses}

    \renewcommand{\arraystretch}{1.15}
    \setlength{\tabcolsep}{6pt}

    \begin{tabular}{lcccc}
        \toprule
        \textbf{QKD link} &
        \textbf{Length} &
        $\boldsymbol{\mathcal{L}_{\mathrm{fib}}}$ &
        $\boldsymbol{\mathcal{L}}$ &
        $\boldsymbol{\mathcal{L}_{\mathrm{comp}}}$ \\
        &
        $(\mathrm{km})$ &
        $(\mathrm{dB})$ &
        $(\mathrm{dB})$ &
        $(\mathrm{dB})$ \\
        \midrule

        N1--N7 & 18.564 & 5.2 & $19.2$ & $14.0$ \\
        
        N7--N6 & 2.347 & 0.8 & $13.0$ & $12.2$ \\
        
        N6--N5 & 5.501 & 1.6 & $11.6$ & $10.0$ \\
        
        N5--N1 & 21.718 & 6.0 & $17.3$ & $11.3$ \\
        
        N2--N4 & 25.765 & 6.5
        & $18.6$ & $12.1$ \\
        
        N4--N3 & 5.045 & 1.9 & $5.1$ & $3.2$ \\
        
        N3--N2 & 21.920 & 5.0 & $16.1$ & $11.1$ \\
        
        N2--N1 & 6.690 & 2.3 & $8.5$ & $6.2$ \\

        \bottomrule
    \end{tabular}
\end{table}

\clearpage
\begin{figure}
    \centering
    \includegraphics[scale=0.47,angle = 90]{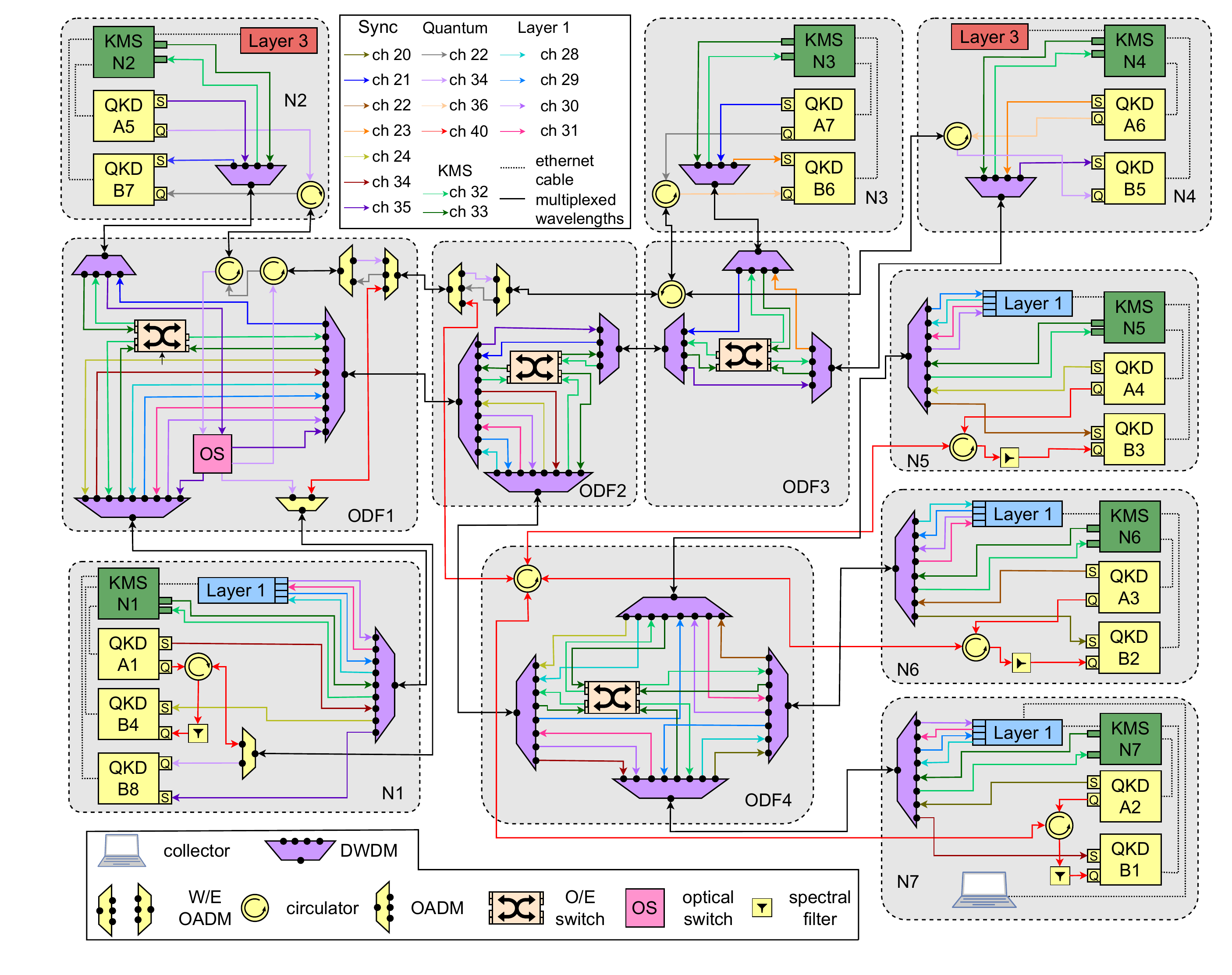}
    \caption{\textbf{Port-level optical architecture of the deployed QKD network.} 
    Quantum, synchronisation, KMS and application-layer connections at nodes N1--N7 and intermediate ODF sites ODF1--ODF4.
    Colours denote the wavelength channel assignments shown in the legend; black dotted lines denote Ethernet connections.
    The optical switch at ODF1 selects the N2--N4 or N2--N1 quantum and synchronisation paths.
    DWDM, dense wavelength division multiplexing filter; W/E OADM, west-to-east optical add/drop module; O/E, optoelectrical.}
    \label{fig:sup:portLevel}
\end{figure}

\clearpage

\section{Supplementary Note 2: QKD protocol and performance model}
The deployed QKD systems implement a time-bin-encoded, three-state QKD protocol with one decoy.
The composable finite-key secret-key-length bound, phase-correlation correction, secrecy and correctness parameters, and privacy-amplification block size are defined in the Methods section of the manuscript.
This Supplementary Note specifies the numerical forward model used to convert the finite-key result into the loss-dependent SKR and QBER predictions shown in Fig.~4a of the main text.

The numerical analysis including the finite-key formulation, detection and error probabilities, and dead-time treatment follow Ref.~\cite{Rusca_2018}.
The error-correction leakage term follows Ref.~\cite{Lim_2014}.
For each quantum channel loss, the numerical model determines time, $T_{\mathrm{block}}$, required to acquire $n_X = 10^6$ clicks in the key generation basis at an emitted pulse repetition rate, $f_{\mathrm{rep}} =$~600~MHz.
The finite-key secret-key rate is then obtained from the secret-key length, eq. (4) of the manuscript, as
\begin{equation}
R_{\mathrm{sec}}^{(0)}
=
\max\!\left(
0,
\frac{\ell_{\mathrm{final}}}{T_{\mathrm{block}}}
\right).
\end{equation}
To account for implementation-specific processing and operational overheads not represented explicitly in the analytical finite-key calculation, the model used an implementation factor, $\eta_{\mathrm{impl}}$, giving
\begin{equation}
R_{\mathrm{sec}}
=
\max\!\left(
0,
\eta_{\mathrm{impl}}
\frac{\ell_{\mathrm{final}}}{T_{\mathrm{block}}}
\right).
\end{equation}
The value $\eta_{\rm impl}=0.30$ was fixed from the vendor and was held constant for all links and channel losses.

The measured total quantum channel loss of each deployed QKD connection was used as the channel-loss input.
Equal preparation probabilities were used for the key-generation and monitoring bases and for the signal and decoy intensity preparation.
The decoy mean photon number was fixed 3~dB below the signal mean photon number, i.e. $\mu_{\rm decoy}=\mu_{\rm signal}/2$, while the signal intensity was optimised independently at each channel loss to maximise the predicted secure key rate.

\begin{table}[H]
\centering
\caption{\textbf{Parameters used in the finite-key performance model.}}
\label{tab:parameters_sim}
\begin{tabular}{lc}
\toprule
Parameter & Value \\
\midrule
Pulse repetition rate, $f_{\mathrm{rep}}$ & $600~\mathrm{MHz}$ \\
Detector efficiency, $\eta_{\mathrm{det}}$ & $0.20$ \\
Total receiver detection efficiency, $\eta_{\mathrm{Bob}}$ & $0.04$ \\
Dark-count probability, $p_{\mathrm{dc}}$ & $8.5\times 10^{-7}$ \\
Detector dead time, $\tau_{\mathrm{det}}$ & $40~\mathrm{\mu s}$ \\
Misalignment error, $e_d$ & $0.01$ \\
Error-correction efficiency, $f_{\mathrm{EC}}$ & $1.14$ \\
Secrecy parameter, $\epsilon_{\mathrm{sec}}$ & $10^{-9}$ \\
Correctness parameter, $\epsilon_{\mathrm{cor}}$ & $10^{-15}$ \\
Privacy-amplification block size, $n_X$ & $10^6$ \\
Phase-correlation parameter, $p_{\mathrm{phc}}$ & $0.02$ \\
Signal/decoy intensity separation & $3~$dB \\
Signal/decoy preparation probability & $1/2$ \\
Key/monitoring-basis preparation probability & $1/2$ \\
Implementation factor, $\eta_{\mathrm{impl}}$ & $0.30$ \\
\bottomrule
\end{tabular}
\end{table}

\section{Supplementary Note 3: Long-term monitoring and data processing}
The long-term performance analysis covers the period from 10 February 2026 at 12:00 to 25 April 2026 at 00:00, corresponding to more than 73 consecutive days of network operation.
SKR, QBER, and further time-tagged generated-key data were obtained from telemetry logs generated by the deployed QKD systems.
For visualisation, SKR measurements were aggregated into one-hour intervals.
QBER measurements were similarly aggregated into one-day intervals for visualisation.
Reported per-link means and standard deviations, however, were calculated from the underlying raw measurements rather than from the aggregated plotting data.
For the optically switched N2--N4 and N2--N1 configurations, only measurements acquired while the corresponding configuration was selected were included in the active-state statistical analysis.

For each QKD-link configuration, the temporal mean and standard deviation of the SKR and QBER were calculated over the corresponding measurement samples.
The network-level values reported as "mean across link configurations" were calculated from the eight per-link temporal means.
The associated uncertainty is the standard deviation across these eight link means, rather than the mean of the individual temporal standard deviations.
The active-state SKR distributions were represented using two-component reflected log-normal mixture fits. The fits provide a descriptive representation of the recurrent higher- and lower-rate operating levels observed for several links and are not used to assign a specific physical origin to either component.

An outage was defined as a contiguous interval longer than $1~\mathrm{h}$ during which a link was scheduled to operate but generated no secret-key material.
Link availability was calculated as the fraction of scheduled operating time outside such outage intervals.
For N2--N4 and N2--N1, the reported values therefore
represent active-state availability.

\section{Supplementary Note 4: Controlled optical switch experiment}

\begin{figure}[h]
    \centering
    \includegraphics[width=1\linewidth]{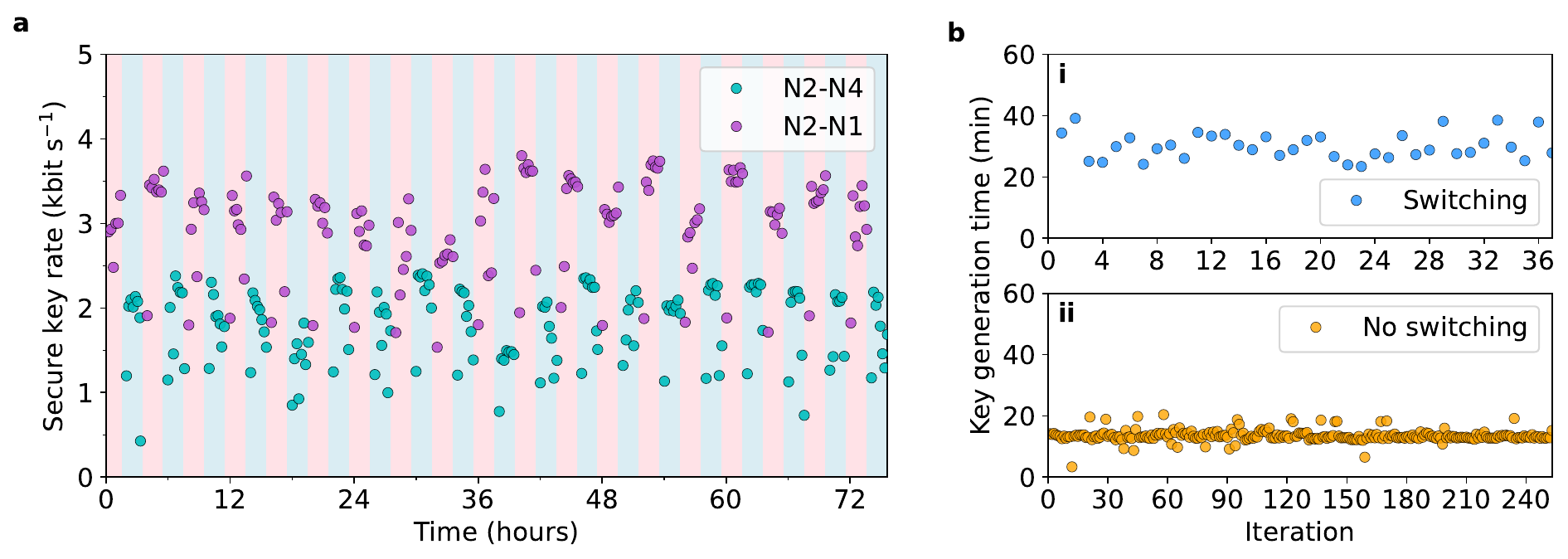}
    \caption{\textbf{Reconfiguration overhead characterisation results.} \textbf{a}, The time series of the secure key rate of the two optically switched QKD links, N2--N4 and N2--N1.
    The shaded regions indicate the period over which each link is active, with light blue corresponding to N2--N4 and pink to N2--N1.
    \textbf{b}, Key generation time. 
    In \textbf{i} we show the case where the links are optically switched and therefore the time includes all the processes taking place after reconfiguration, in addition to the key generation time.
    In \textbf{ii}, only static link events are plotted as a baseline for characterising the reconfiguration overhead.}
\label{fig:optical_switch_experiment}
\end{figure}

During the 73-day monitoring period, we performed a controlled optical switching experiment in order to characterise the additional latency required to reconfigure the network. 
In this experiment, the N2--N4 and N2--N1 configurations were alternated every $2~\mathrm{h}$ for approximately three days.
Supplementary Fig.~\ref{fig:optical_switch_experiment}a shows a zoomed view of the secret key rate traces during the controlled switching experiment, which took place on days 35--37 of the monitoring period.
In Fig.~\ref{fig:optical_switch_experiment}b we present the results by comparing the time required to generate a key when the link is switched, \textbf{i} and when the link is static, \textbf{ii}.
In the primer, the elapsed time includes synchronisation, calibration and key generation.

Across 37 switching events, the average time to switch and generate a key was $30\pm4$ min, whereas for 253 static link intervals, the corresponding key generation time was $13\pm2$ min.
By subtracting the static link key generation time from the time required to generate a key in the switched samples, the additional delay attributed to the optical switch reconfiguration was determined to be $17\pm5$ min.
This difference is therefore defined as the reconfiguration overhead, representing the additional time required for optical switching, link calibration, optimisation, and start-up.

\section{Supplementary Note 5: End-to-end key delivery and route reconstruction}
\begin{figure}[H]
    \centering
    \includegraphics[width=1\linewidth]{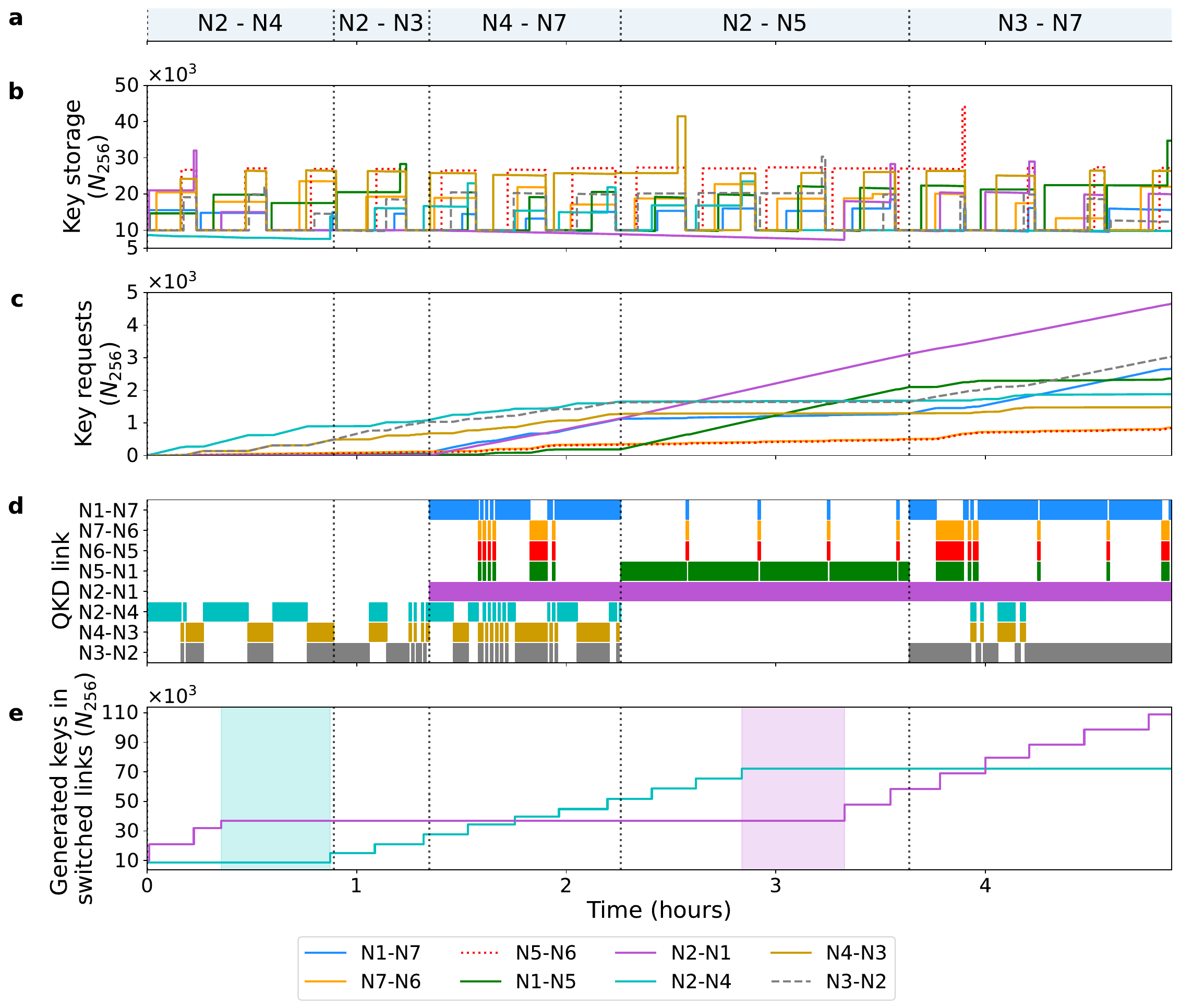}
    \caption{\textbf{End-to-end key request experiment.}
    \textbf{a}, KMS source--destination schedule used during the experiment.
    \textbf{b}, Stored-key level associated with each physical QKD link.
    \textbf{c}, Cumulative request-associated key consumption attributed to each physical QKD link.
    \textbf{d}, Reconstructed QKD-link involvement for the delivered end-to-end keys.
    Coloured segments indicate physical QKD links belonging to topology-compatible direct or trusted-relay paths reconstructed from the processed link-pool data. 
    \textbf{e}, Generated key material on the optically switched N2--N4 and N2--N1 links.
    Shaded regions indicate the transition interval between optical-switch reconfiguration and the onset of fresh key generation on the selected link, including synchronisation recovery, calibration and key-generation.}
    \label{fig:sup:e2e}
\end{figure}

The end-to-end experiment was performed through the software-defined network (SDN) controller located at N7 and communicating with the KMS interface.
Five source--destination pairs were sequentially selected over approximately $5~\mathrm{h}$: N2--N4, N2--N3, N4--N7, N2--N5, and N3--N7, Supplementary Fig.~\ref{fig:sup:e2e}a.
For each request interval, the source--destination pair and target key volume were selected at random, and sequential 256-bit end-to-end keys were requested until the assigned target volume was delivered, before proceeding to the next interval.
The Layer~1 and Layer~3 encryption services remained active throughout the experiment and continued to consume key material from their corresponding link-local key pools.
For each end-to-end request, the controller selected a direct or trusted-relay path using the proprietary Dijkstra routing algorithm of the vendor.

Active QKD links continuously replenished their corresponding link-local key pools.
In the commercial devices, every $20~\mathrm{min}$, older keys are automatically discarded to accommodate newly generated key material in order to maintain a predefined storage maximum of $10000$ keys.
These automatic discard events are identified as an instant drop to $10000$ keys in the corresponding key storage, see Supplementary Fig.~\ref{fig:sup:e2e}b.

The direct or trusted relay path, as chosen by the algorithm, results in key requests in the selected links, Supplementary Fig.~\ref{fig:sup:e2e}c.
The different slopes reflect variations in the rate of key consumption associated with successive end-to-end key requests.
The physical QKD links involved in the selected route for fulfilling each end-to-end key request are indicated by the shaded regions in Supplementary Fig.~\ref{fig:sup:e2e}d.

During end-to-end key requests, the stored key pools of the optically switched links were progressively depleted via the KMS layer.
When depletion caused the stored keys in either of the optically switched links to drop below 8000, the controller reconfigured the optical switch in the quantum layer to activate the corresponding QKD link in order to replenish its storage.
The secure key generation of both switched links throughout the experiment is shown in Supplementary Fig.~\ref{fig:sup:e2e}e, where intervals without key generation correspond to periods during which the respective link was inactive.
Optical reconfiguration occurred during two of the five request intervals; following each switch, the newly selected QKD link underwent synchronisation and calibration before fresh key generation resumed (shaded areas).
During these intervals, previously accumulated link-local key material remained available to the KMS and could continue to support end-to-end delivery, provided that sufficient stored key was available along the selected route.

\section*{References}

\begin{enumerate}
\item D. Rusca, A. Boaron, F. Grünenfelder, A. Martin, and H. Zbinden,
Finite-key analysis for the 1-decoy state QKD protocol.
\textit{Applied Physics Letters} \textbf{112}, 171104 (2018).

\item C. C. W. Lim, M. Curty, N. Walenta, F. Xu, and H. Zbinden,
Concise security bounds for practical decoy-state quantum key distribution.
\textit{Physical Review A} \textbf{89}, 022307 (2014).
\end{enumerate}

\end{document}